\documentclass[prd,preprint,tightenlines,showpacs,preprintnumbers,nofootinbib,eqsecnum,superscriptaddress]{revtex4}

 \usepackage[dvips,final]{graphicx}
  \usepackage{amssymb}
   \usepackage{amsmath}
    \usepackage{amsfonts}
     \usepackage{bm}

\begin{document}

\title{Inclusive production of Higgs boson 
in the two-photon channel \\at the LHC 
within $\bm{k_{t}}$-factorization approach 
and with\\ the Standard Model couplings
} 

\author{Antoni~Szczurek}
\email{antoni.szczurek@ifj.edu.pl}
\affiliation{Institute of Nuclear Physics PAN, PL-31-342 Krak\'ow, Poland}
\affiliation{University of Rzesz\'ow, PL-35-959 Rzesz\'ow, Poland}

\author{Marta~{\L}uszczak}
\email{luszczak@univ.rzeszow.pl} \affiliation{University of Rzesz\'ow, PL-35-959 Rzesz\'ow, Poland}

\author{Rafa{\l}~Maciu{\l}a}
\email{rafal.maciula@ifj.edu.pl} \affiliation{Institute of Nuclear Physics PAN, PL-31-342 Krak\'ow, Poland}

\date{\today}

\begin{abstract}
We calculate differential cross sections for Higgs boson and/or 
two-photon production from intermediate (virtual) Higgs boson within
the formalism of $k_t$-factorization. The off-shell $g^* g^* \to H$ 
matrix elements are used. We compare results obtained 
with infinite top fermion (quark) mass and with finite mass 
taken into account. The latter effect is rather small.
We compare results with different unintegrated gluon distributions 
from the literature. Two methods are used. In the first method first
Higgs boson is produced in the $2 \to 1$ $g g \to H$ $k_t$-factorization
approach and then isotropic decay with the Standard Model branching 
fraction is performed. In the second method we calculate directly
two photons coupled to the virtual Higgs boson. The results of the 
two methods are compared and differences are discussed.
The results for two photons from the Higgs boson are compared with 
recent ATLAS collaboration data.
In contrast to a recent calculation the leading order $g g \to H$ 
contribution is rather small compared to the ATLAS experimental data
($\gamma \gamma$ transverse momentum and rapidity distributions)
for all unintegrated gluon distributions from the literature.
We include also higher-order contribution 
$g g \to H (\to \gamma \gamma) g$, $g g \to g H g$ and the contribution 
of the $W^+ W^-$ and $Z^0 Z^0$. 
The $gg\to Hg$ mechanism gives similar cross section as the $gg\to H$ 
mechanism.
We argue that there is almost no double counting when adding 
$gg\to H$ and $gg\to Hg$ contributions due to different topology of Feynman diagrams.
The final sum is comparable with the ATLAS two-photon data. 
We discuss uncertainties related to both the theoretical approach and existing UGDFs. 
\end{abstract}



\maketitle

\section{Introduction}

The Higgs-like boson has been discovered recently at the LHC
\cite{Higgs_discovery}. It has been observed in a few decay channels.
The $\gamma \gamma$ and $Z^0 Z^{0,*}$ are particularly spectacular
\cite{Aad:2013wqa,Chatrchyan:2013mxa,Khachatryan:2014iha,Aad:2014aba}.
Before the discovery many of the Higgs properties were strongly
dependent on its mass. Now knowing the Higgs boson mass 
$M_H \approx$ 126 GeV
we can fix parameters for production and decay of the Higgs boson,
at least within the Standard Model.
We slowly enter era of more detailed studies.
In particular, it is very important to know what is the Higgs boson spin 
and parity and if it is Standard Model object.
Also understanding the rapidity and transverse momentum distributions
is particularly interesting. While the total cross section is well
under control and was calculated in leading-order (LO), next-to-leading order
(NLO) and even next-to-next-to-leading order (NNLO) approximation \cite{NNLO} 
the distribution in the Higgs boson transverse
momentum is more chalanging. This can be addressed e.g. in transverse
momentum resummation approach (see e.g. Refs.~\cite{FFGT2011,FFGT2012} 
and references therein).
 
It was advocated recently that precise differential data for Higgs boson
in the two-photon final channel could be very useful to test and explore
unintegrated gluon distribution functions (UGDFs) \cite{Jung2013}. 
It was shown very recently \cite{LMZ2014} that the $k_t$-factorization 
formalism with commonly used UGDFs (Kimber-Martin-Ryskin (KMR)
\cite{KMR} and Jung CCFM \cite{Jung}) gives a reasonable description of 
recent ATLAS data obtained at $\sqrt{s} =$ 8 TeV \cite{ATLAS_Higgs}. 
We perform similar calculation and, as will be seen in the following, 
draw rather different conclusions.

In the present study we present several differential distributions
for the Higgs boson and photons from the Higgs boson decay
at $\sqrt{s}$ = 8 TeV for various UGDFs from the literature, also the 
ones used in the context of low-$x$ physics (Kutak-Sta{\'s}to 
\cite{Kutak-Stasto} and Kutak-Sapeta \cite{Kutak-Sapeta}). 
We include both leading-order and next-to-leading order contributions.
We shall critically discuss uncertainties and open problems
in view of the recent ATLAS data.

\section{Formalism}

\subsection{Higgs boson production}

\begin{figure*}
\begin{center}
\includegraphics[width=8cm]{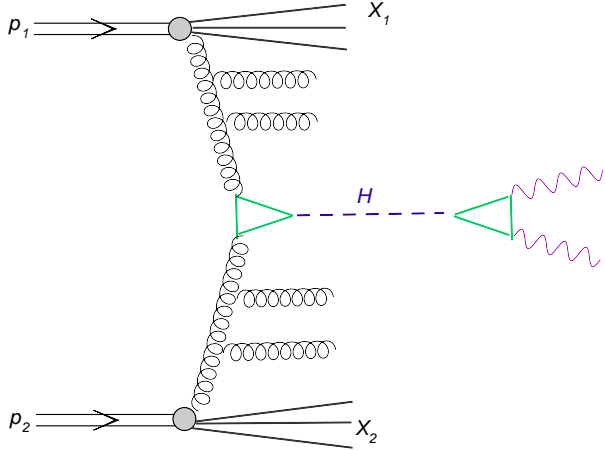}
\caption{Dominant leading-order diagram for inclusive Higgs boson
production in the two-photon channel.
}
\label{fig:gg_H}
\end{center}
\end{figure*}

In the $k_t$-factorization approach the cross section for 
the Higgs boson production can be written somewhat formally as:
\begin{eqnarray}
\sigma_{pp \to H} = \int \frac{dx_1}{x_1} \frac{dx_2}{x_2}
\frac{d^2 q_{1t}}{\pi} \frac{d^2 q_{2t}}{\pi} 
&&\delta \left( (q_1 + q_2)^2 - M_H^2 \right) 
\sigma_{gg \to H}(x_1,x_2,q_{1},q_{2}) \nonumber \\
&&\times \; {\cal F}_g(x_1,q_{1t}^2,\mu_F^2) {\cal F}_g(x_2,q_{2t}^2,\mu_F^2)
\; ,
\label{Higgs_kt_factorization}
\end{eqnarray}
where ${\cal F}_g$ are so-called unintegrated (or 
transverse-momentum-dependent) gluon distributions
and $\sigma_{g g \to H}$ is $g g \to H$ (off-shell) cross section.
The situation is illustrated diagramatically in Fig.~\ref{fig:gg_H}.
%

It is easy to show in the collinear approximation (see e.g. 
Ref.~\cite{ESW_book}) that:
\begin{equation}
\sigma_{gg \to H} = \frac{\pi}{\hat s} \delta(\hat s - m_H^2) \; .
\label{cross_section}
\end{equation}
After some manipulation the formula (\ref{Higgs_kt_factorization}) 
can be written conveniently as (see Ref.~\cite{LS2006_Higgs})
\begin{eqnarray}
\sigma_{pp \to H} = \int d y d^2 p_t d^2 q_t \frac{1}{s x_1 x_2}
\frac{1}{m_{t,H}^2}
\overline{|{\cal M}_{g^*g^* \to H}|^2} 
{\cal F}_g(x_1,q_{1t}^2,\mu_F^2) {\cal F}_g(x_2,q_{2t}^2,\mu_F^2) / 4
\; ,
\label{useful_formula}
\end{eqnarray}
which can be also used to calculate rapidity and transverse
momentum distribution of the Higgs boson.

In the last equation:
$\vec{p}_t = \vec{q}_{1t} + \vec{q}_{2t}$ is transverse momentum 
of the Higgs boson
and $\vec{q}_t = \vec{q}_{1t} - \vec{q}_{2t}$ is auxiliary variable 
which is used in the integration. Furthermore:
$m_{t,H}$ is the so-called Higgs transverse mass and
$x_1 = \frac{m_{t,H}}{\sqrt{s}} \exp( y)$,  
$x_2 = \frac{m_{t,H}}{\sqrt{s}} \exp(-y)$.
The factor $\frac{1}{4}$ is the jacobian of transformation from
$(\vec{q}_{1t}, \vec{q}_{2t})$ to $(\vec{p}_t, \vec{q}_{t})$ variables.

Similar formalism was used in the past for production of gauge bosons
\cite{KS2004}. There gluon distributions have to be replaced by
unintegrated quark and antiquark distributions.

Let us concentrate for a while on the matrix element for the 
$g^* g^* \to H$. In Ref.~\cite{LS2006_Higgs} the on-shell matrix element was
used. In Ref.~\cite{LZ2005} the off-shell matrix element was used instead,
however in the approximation of infinitly heavy top in 
the triangle-coupling of gluons to the Higgs boson
(see also\cite{Hautmann:2002tu} where the off-shell matrix element 
was discussed).
Then the effective $g^* g^* \to H$ coupling is relatively simple. 
The matrix element under discussion (for on-shell Higgs boson) 
takes the simple form: 
\begin{equation}
{\cal M}_{g^* g^* \to H}^{ab} = -i \delta^{a b} \frac{\alpha_s}{4 \pi} 
\frac{1}{v} 
\left( m_H^2 + p_t^2 \right) cos (\phi) \frac{2}{3} \; ,
\label{gg_H_approximate_ME}
\end{equation}
where $v^2 = \left( G_F \sqrt{2} \right)^{-1}$.
The effect of finite-mass corrections was studied in Ref.~\cite{PTS2006}
in the context of $k_t$-factorization and in \cite{Marzani:2008az}
in the context of higher-order collinear approximation corrections.
Then the corresponding matrix element is more complicated
and can be written with the help of two form factors:
\begin{equation}
{\cal M}_{g^* g^* \to H}^{ab} = -i \delta^{a b} \frac{\alpha_s}{4 \pi} 
\frac{1}{v} \left[ \left( m_H^2 + p_t^2 \right) cos (\phi) G_1(q_1, q_2, q)
- \frac{ 2 (m_H^2 + p_t^2)^2 |q_{1t}|
  |q_{2t}|}{(m_H^2+q_{1t}^2+q_{2t}^2)} G_2(q_1, q_2, q) \right]
 \; .
\label{better_ME}
\end{equation}
The form factors $G_1$ and $G_2$ have an integral representation.
However, at not too big virtualities of gluons and Higgs boson 
the following approximate formula for the $G_1$ and $G_2$ form factors 
can be used \cite{PTS2006}:
\begin{eqnarray}
G_1 &=& \frac{2}{3} \left( 1 + \frac{7}{30} \chi + \frac{2}{21} \chi^2 +
  \frac{11}{30}(\xi_1 + \xi_2) + ... \right) \; ,  \\
G_2 &=& -\frac{1}{45} (\chi - \xi_1 - \xi_2) - \frac{4}{315} \chi^2 +
.... \; ,
\label{G1_G2}
\end{eqnarray}
where the expansion variables $\chi$, $\xi_1$, $\xi_2$ above are defined as:
\begin{eqnarray}
\chi  &=& \frac{q^2}{4 m_f^2} \; , \\
\xi_1 &=& \frac{q_1^2}{4 m_f^2} < 0 \; ,\\
\xi_2 &=& \frac{q_2^2}{4 m_f^2} < 0 \; .
\label{expansion_ratios}
\end{eqnarray}
%

\subsection{$H \to \gamma \gamma$}

The matrix element for the Higgs boson decay into photons with
helicity $\lambda_1$ and $\lambda_2$ can be written as
\begin{equation}
{\cal M}_{H \to \gamma \gamma}(\lambda_1, \lambda_2) =
T_{H \to \gamma \gamma}^{\mu \nu} 
\epsilon_{\mu}^*(\lambda_1)
\epsilon_{\nu}^*(\lambda_2) \; .
\label{ME_H_gamgam}
\end{equation}
The leading-order (LO) vertex function can be decomposed as the sum
\begin{equation}
T_{H \to \gamma \gamma}^{\mu \nu} = 
T_{H \to \gamma \gamma}^{\mu \nu, W} +
T_{H \to \gamma \gamma}^{\mu \nu, t} + ... \; ,
\label{vertex_H_gamgam}
\end{equation}
where the first term includes loops with intermediate  $W^{\pm}$
and the second term triangle(s) with top quarks. The dots
represent contribution of triangles with bottom and charm quarks
and with $\tau$ leptons, etc.
The vertex function can be written as:
\begin{equation}
T_{H \to \gamma \gamma}^{\mu \nu}(p_1,p_2) =
i \frac{\alpha_{em}}{2 \pi} {\cal A}
\left( G_F \sqrt{2} \right)^{1/2}
\left(p_2^{\mu} p_1^{\nu} - (p_1 \cdot p_2) g^{\mu \nu} \right) \; .
\label{vertex_function_H_gamma_gamma}
\end{equation}
In the Standard Model the $\cal A$ constant is:
\begin{equation}
{\cal A} = {\cal A}_W(\tau_W) + N_c e_f^2 {\cal A}_t(\tau_t) + ...
\label{cal_A}
\end{equation}
where the arguments are:
\begin{equation}
\tau_W = \frac{m_H^2}{4 m_W^2} \;\;\;\; , \;\;\;\; 
\tau_t = \frac{m_H^2}{4 m_t^2} \; .
\end{equation}
The functions ${\cal A}_W$ and ${\cal A}_t$ have the simple form:
\begin{eqnarray}
{\cal A}_W(\tau) &=& 
-\left( 2 \tau^2 + 3 \tau + 3(2 \tau-1)f(\tau) \right) /\tau^2
\; ,\\
{\cal A}_t(\tau) &=& 
2 \left( \tau + (\tau - 1)f(\tau) \right) / \tau^2 \; ,
\end{eqnarray}
where the function $f(\tau)$ reads:
\begin{equation}
f(\tau) = arc sin^2(\sqrt{t}) \; .
\end{equation}
For light fermions the function $f(\tau)$ is slightly different \cite{Barger:1987nn}.

The two-photon decay width can be calculated as:
\begin{equation}
\Gamma_{H \to \gamma \gamma} = \frac{1}{32 \pi^2}  
\Sigma_{\lambda_1 \lambda_2}
|{\cal M}_{H \to \gamma \gamma}(\lambda_1,\lambda_2)|^2  
\frac{p}{m_H^2} \frac{1}{2} \; .
\label{decay_width}
\end{equation}
The factor $\frac{1}{2}$ is due to identity of the final state photons.
Using Eq.(\ref{decay_width}) with matrix element given by
Eq.(\ref{ME_H_gamgam}) we get 
$\Gamma_{H \to \gamma \gamma}$ = 0.91 $\times$ 10$^{-5}$
which, when combined with the total decay width 
$\Gamma_H \approx$ 4 MeV \cite{DHPRS2011}, gives branching fraction
$\textrm{BF}_{H \to \gamma \gamma}$ = 2.27 $\times$ 10$^{-3}$, consistent
with what is known from the literature (see e.g. Ref.~\cite{DKS98}).
Using the decay matrix element from Ref.~\cite{LMZ2014}
would give much bigger $\textrm{BF}_{H \to \gamma \gamma} \sim$ 0.01 
(incorrect) branching fraction. Two-loop corrections are rather 
very small \cite{LL1997}.

\subsection{$g^* g^* \to H^* \to \gamma \gamma$}

Let us combine now all elements defined above and write matrix element
for the $g^* g^* \to H^* \to \gamma \gamma$ process.
\begin{equation}
{\cal M}_{g^{*} g^{*} \to H^{*} \to \gamma \gamma}(\lambda_1, \lambda_2) =
{\cal M}_{g^{*}g^{*} \to H^{*}}(\vec{q}_{1t},\vec{q}_{2t};\hat s)
\frac{1}{{\hat s} - M_H^2 + i \Gamma_H M_H} 
{\cal M}_{H^{*} \to \gamma \gamma}(\lambda_1, \lambda_2) \; .
\label{gg_H_gamgam}
\end{equation}
In the infinitly heavy quark approximation the matrix element squared 
averaged over colors can be written in the quite compact way 
(see Ref.~\cite{LMZ2014}):
\begin{equation}
\overline { | {\cal M} |^2 } = \frac{1}{1152 \pi^4}
\alpha_{em}^2 \alpha_s^2 G_F^2 | {\cal A} |^2
\frac{ {\hat s}^2 (\hat s + p_t^2)^2 }{( \hat s - m_H^2)^2 + m_H^2
  \Gamma_H^2}     \cos^2(\phi) \; .       
\label{ME_gg_H_gamgam}
\end{equation}
The differential (in photon rapidities $y_1$, $y_2$ and 
transverse momenta $p_{1t}$, $p_{2t}$) 
cross section for the production of a pair of photons 
from the $g^* g^* \to H^* \to \gamma \gamma$ subprocess with intermediate 
virtual Higgs boson can be written as:
\begin{eqnarray}
\frac{d \sigma(p p \to H X \to \gamma \gamma X)}
{d y_1 d y_2 d^2 p_{1t} d^2 p_{2t}} 
&& = \frac{1}{16 \pi^2 {\hat s}^2} \cdot \frac{1}{2} \cdot \int \frac{d^2 k_{1t}}{\pi} \frac{d^2
  k_{2t}}{\pi} \overline{|{\cal M}^{off}_{g^{*}g^{*} \to H^{*} \to
\gamma \gamma}|^2} \nonumber \\
&& \!\!\!\!\!\!\!\!\!\!\!\!\!\!\!\!\!\!  \times \;\; \delta^2 \left( \vec{k}_{1t} + \vec{k}_{2t} - \vec{p}_{1t} - \vec{p}_{2t}
\right)
{\cal F}_g(x_1,k_{1t}^2,\mu^2) {\cal F}_g(x_2,k_{2t}^2,\mu^2) \; .
\label{kt_2_to_2}
\end{eqnarray}
Please note that in this case the $m_H^2 + p_t^2$ term in 
Eq.(\ref{gg_H_approximate_ME}) for on shell Higgs boson is replaced 
by $ \hat s + p_t^2$ for virtual Higgs boson.
This has consequences some distance from the resonance
position where the cross section is however small.
In principle, also $M_H^2$ in definition of the $\cal A$ functions should
be replaced by $\hat s$ here.

Since we integrate over full phase space in $y_1$, $y_2$, 
$p_{1t}$ and $p_{2t}$ we have to include in addition identity
factor $\frac{1}{2}$, in full analogy to the calculation
of the decay width into two photons.

How to remove the $\delta$ function in Eq.(\ref{kt_2_to_2}) in a convenient 
for calculation way is described in Ref.~\cite{LS2006_ccbar}.
The calculation of the cross section according to formula
(\ref{kt_2_to_2}) with matrix element (\ref{ME_gg_H_gamgam})
is not easy as the light Higgs boson discovered recently is a very
narrow resonance.
This calculation is performed within a Monte Carlo method using
a well know package VEGAS \cite{Lepage1978}.
We have carefully tested both numerics and convergence.

\subsection{$g g \to H g$}

\begin{figure*}
\begin{center}
\includegraphics[width=6cm]{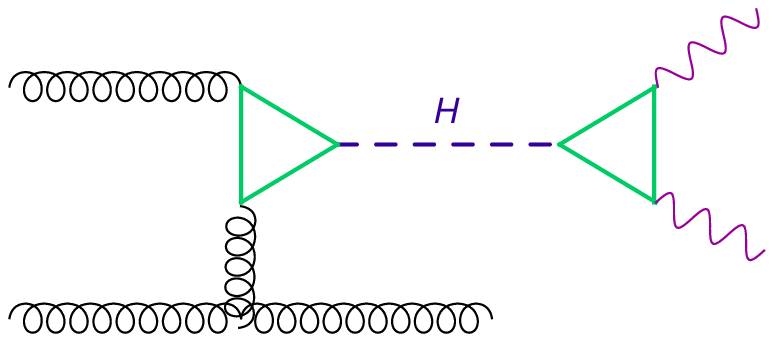}
\includegraphics[width=6cm]{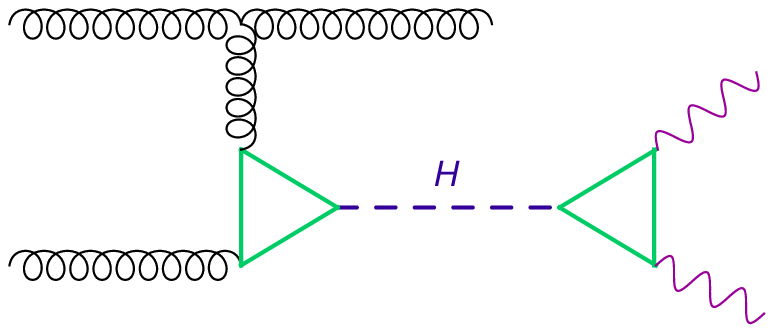}
\includegraphics[width=6cm]{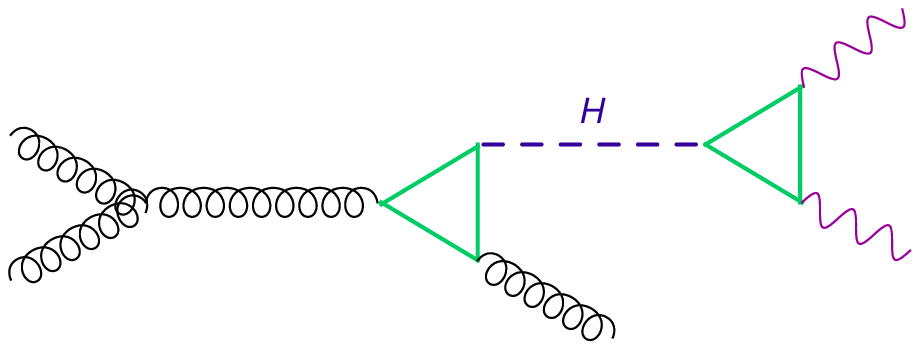}\\
\includegraphics[width=6cm]{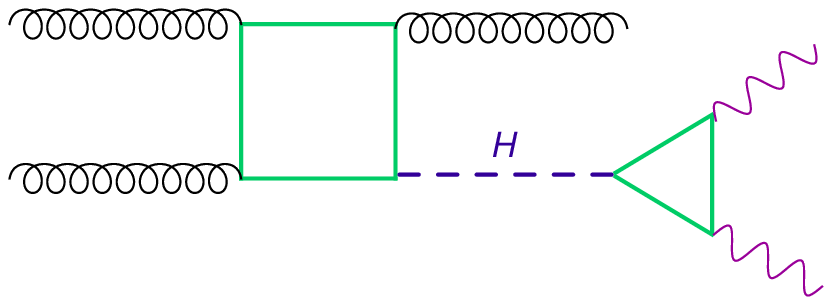}
\includegraphics[width=6cm]{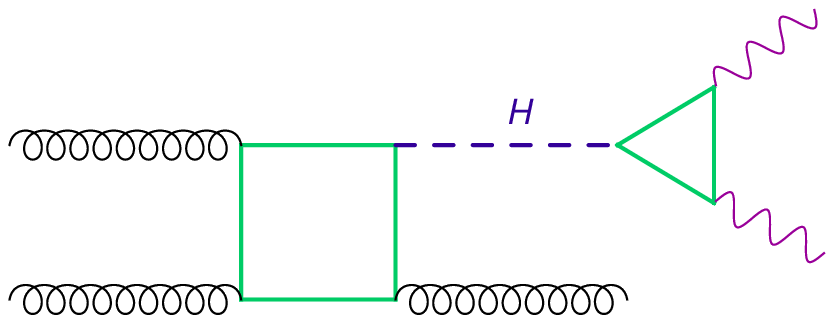}
\caption{Typical diagrams for QCD NLO contributions 
to the Higgs boson production.
}
\label{fig:NLO}
\end{center}
\end{figure*}

In the collinear-approximation the cross section for fixed-order
processes of the type $p_1 p_2 \to H p_3$ (parton1$+$parton2 $\to$
Higgs$+$parton3) (see Fig.~\ref{fig:NLO})
of the order of $\alpha_s$ is well known
since long time \cite{2to2_basis}.

The corresponding cross section differential in Higgs boson rapidity ($y_H$),
associated parton rapidity ($y_p$) and transverse momentum of each
of them can be written as:
\begin{eqnarray}
\frac{d \sigma}{dy_H dy_p d^2 p_t}(y_H,y_p,p_t) &=&
\frac{1}{16 \pi^2 {\hat s}^2 }  
\times \biggl\{ x_1 g_1(x_1,\mu^2)  x_2 g_2(x_2,\mu^2) 
 \overline{|{\cal M}_{gg \to Hg}|^2}   \nonumber \\   
&&+ \left[ \sum_{{f_1}=-3,3} x_1 q_{1,f_1}(x_1,\mu^2) \right] 
x_2 g_2(x_2,\mu^2)  \;
 \overline{|{\cal M}_{qg \to Hq}|^2}   \nonumber \\ 
&& + \; x_1 g_1(x_1,\mu^2)  
 \displaystyle \left[ \sum_{{f_2}=-3,3} x_2 q_{2,f_2}(x_2,\mu^2) \right]       
 \overline{|{\cal M}_{gq \to Hq}|^2}   \nonumber \\ 
&&+ \sum_{f=-3,3} x_1 q_{1,f}(x_1,\mu^2)  x_2 q_{2,-f}(x_2,\mu^2)
 \overline{|{\cal M}_{qq \to Hg}|^2}  \biggr\} .   
\label{2to2}
\end{eqnarray}
The indices $f$ in the formula above number both quarks ($f >$ 0)
and antiquarks ($f <$ 0). Only three light flavours are included in
actual calculations here.
The explicit formulae for $\overline{|{\cal M}|^{2}}$ can be found
in Ref.~\cite{2to2_basis}.
We have checked that the $g g \to H g$ contribution dominates
over the two other types of contributions. This can be understood
as due to the presence of the box contributions for $g g \to H g$ 
but absent in the other cases.

In the following we shall calculate the dominant $g g \to H g$
contribution also taking into account transverse momenta of initial gluons.
In the $k_t$-factorization the NLO differential cross section can 
be written as:
\begin{eqnarray}
\frac{d \sigma(p p \to H g X)}{d y_H d y_g d^2 p_{H,t} d^2 p_{g,t}}
&& = 
\frac{1}{16 \pi^2 {\hat s}^2} \int \frac{d^2 q_{1t}}{\pi} \frac{d^2 q_{2t}}{\pi} 
\overline{|{\cal M}_{g^{*} g^{*} \rightarrow H g}^{off-shell}|^2} 
\nonumber \\
&& \times \;\; 
\delta^2 \left( \vec{q}_{1t} + \vec{q}_{2t} - \vec{p}_{H,t} - \vec{p}_{g,t} \right)
{\cal F}(x_1,q_{1t}^2,\mu^2) {\cal F}(x_2,q_{2t}^2,\mu^2) \; .
\label{kt_fact_gg_Hg}
\end{eqnarray}
This can be further simplified as discussed e.g. in Ref.~\cite{LS2006_ccbar}.

Calculation of the off-shell matrix element for the process under
consideration is rather complicated in the most general case as 
it involves loops (triangles and boxes).
Since the box diagrams with very heavy top quarks/antiquarks
dominate at high energies we expect that the off-shell 
effects should be relatively small.
In the present approach we make the following replacement to simplify
the calculation:
\begin{equation}
\overline{|{\cal M}_{g^*g^* \to Hg}^{off-shell}|^2} \rightarrow
\overline{|{\cal M}_{gg \to Hg}^{on-shell}(s,t,u)|^2} \; ,
\end{equation}
where the latter is analytical continuation of the on-shell
matrix element off mass shell. The larger $q_{1t}$ or $q_{2t}$
the worse the approximation could be. This cannot be
quantified, however, before exact off-shell matrix element is calculated.
This goes beyond the scope of the present study.

\subsection{Higgs boson and dijets in the context of
$k_t$-factorization approach}

It is well known that in contrast to gauge boson ($W^{\pm}$ and $Z^0$) 
production for calculating inclusive cross section for the Higgs boson 
production not only LO but also NLO and even NNLO corrections are pretty large.
Collinear NNLO contributions to the Higgs boson production associated
with dijet production was discussed e.g. in Ref.~\cite{Duca_Hjj}. 
A somewhat simplified but pedagogical high-energy approach was discussed 
in Ref.~\cite{Duca_high_energy}.

\begin{figure*}
\begin{center}
\includegraphics[width=4cm]{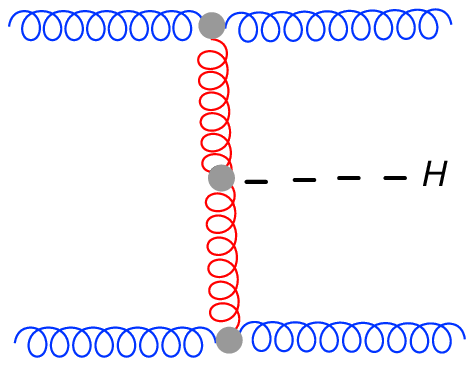}
\includegraphics[width=4cm]{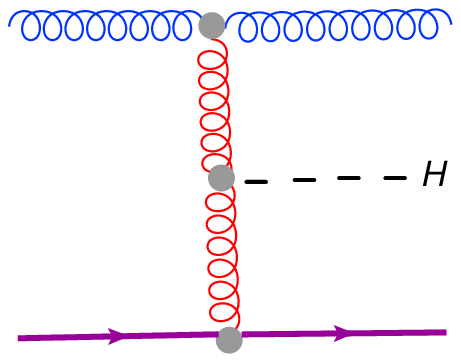}
\includegraphics[width=4cm]{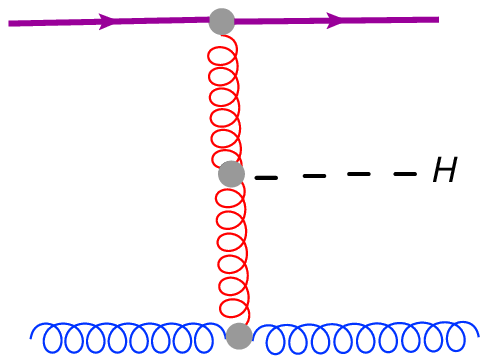}
\includegraphics[width=4cm]{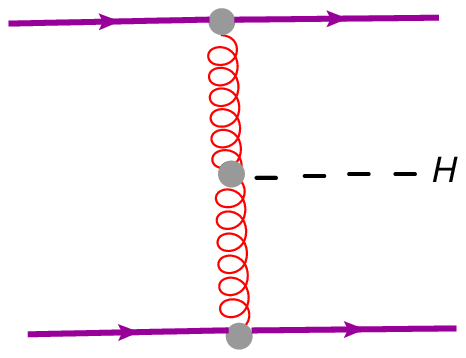}
\caption{The $2 \to 3$ diagrams which are used in order to make
reference to the $2 \to 1$ $k_t$-factorization calculation.
}
\label{fig:gg_gHg}
\end{center}
\end{figure*}

In the present analysis we wish to make a reference to 
the $g g \to H$ $k_t$-factorization calculations
so a simplified approach may be useful.
In the following we shall evaluate cross section and differential 
distributions in the collinear approximation for the subprocesses
shown in Fig.~\ref{fig:gg_gHg}. At large $q_{1t}$ and $q_{2t}$
(transverse momenta of the exchanged (red online) gluons)
the contribution of the first subprocess
($g g \to g H g$) can be directly compared to the $k_t$-factorization
result with the KMR UGDF. This may be useful in order to understand
higher-order contributions contained in the $k_t$-factorization approach.

The matrix element for the $g g \to g H g$ which can (should) be used 
to compare the collinear factorization result with the
$k_t$-factorization approach result reads:
\begin{eqnarray}
{\cal M}_{\lambda_1 \lambda_2 \to \lambda_3 \lambda_4}^{g g \to g H g}
(ace,bde') = &&
g_s(\mu_{r,1}^2) f_{ace} \epsilon_{\mu_1}(\lambda_1) 
C^{\mu_1 \nu_1 \tau_1}(-p_1,p_3,q_1) \epsilon_{\nu_1}^*(\lambda_3)
\nonumber \\
&&
\frac{(-ig_{\tau_1 \tau_1'})}{t_1} 
T_{g g \to H}^{\tau_1' \tau_2'}(q_1,q_2,p_H)
\frac{(-ig_{\tau_2 \tau_2'})}{t_2}
\nonumber \\
&&
g_s(\mu_{r,2}^2) f_{bde'} \epsilon_{\mu_2}(\lambda_2)
C^{\mu_2 \nu_2 \tau_2}(-p_2,p_4,q_2) \epsilon_{\nu_2}^*(\lambda_4)
\; ,
\label{gg_gHg_full}
\end{eqnarray}
where dependence on renormalization scale was made explicit.

Here the matrix element is evaluated as in Ref.~\cite{Duca_high_energy} 
using high-energy approximations.
It can be written somewhat schematically as:
\begin{equation}
\overline{|{\cal M}_{g g \to gHg}|^2} =
4 \frac{C_A^2}{N_c^2 - 1} g_s^2(\mu_{r,1}^2) g_s^2(\mu_{r,2}^2)
\frac{\hat s}{t_1^2 t_2^2} |C_{gg \to H}(q_1,p_H,q_2)|^2 \; .
\label{gg_gHg}
\end{equation}
The matrix element is particularly simple in the limit:
\begin{equation}
s_{ij} \gg s_{iH},s_{jH} \gg m_H^2 \; .
\label{special_limit}
\end{equation}
We have made explicit running of strong coupling constant
in (\ref{gg_gHg}).
In practical calculation it is reasonable to take 
$\mu_{r,1}^2 = p_{3t}^2$ and $\mu_{r,2}^2 = p_{4t}^2$.
At high energies $t_1 \approx -q_{1t}^2 = -p_{3t}^2$ and 
$t_2 \approx -q_{2t}^2 = -p_{4t}^2$.

The phase space integration is performed then with the $g g \to g H g$
matrix element squared and collinear gluon distribution functions (GDFs), 
see for example next subsection. 
Both integrated and differential cross sections can be then compared
with those obtained within the $k_t$-factorization approach.
Especially inspiring is to understand the interrelation between the
two approaches for larger jet/Higgs transverse momenta 
$p_{3t}, p_{4t}$, $p_{tH}$.

In the high-energy approach quark and antiquarks contributions can 
be easily included by replacing gluon distributions 
$g(x_1,\mu_{f,1}^2)$ and $g(x_2,\mu_{f,2}^2)$ 
by so-called effective parton distributions (see e.g. Ref.~\cite{BP_book}):
\begin{eqnarray}
f_{eff}(x_k,\mu_k^2) &=& g(x_k,\mu^2) \nonumber \\
&+& \frac{C_F}{C_A}
\left(  u(x_k,\mu_k^2) + d(x_k,\mu_k^2) + s(x_k,\mu_k^2)
      + {\bar u}(x_k,\mu_k^2) + {\bar d}(x_k,\mu_k^2) + {\bar s}(x_k,\mu_k^2)
\right) \; . \nonumber \\
\label{effective_PDF}
\end{eqnarray}
Similar procedure is often done in the context of Mueller-Navelet jets.
We shall evaluate and show the quark/antiquark components separately 
as they are not taken into account explicitly in 
the $k_t$-factorization approach.

\subsection{$\bm{WW}$ fusion}

Now we wish to consider purely electroweak corrections
that are known to give sizeable contribution to the Higgs boson production. 

The second most important mechanism for the Higgs boson production
is the fusion of off-shell gauge bosons: $WW$ or $ZZ$. It is known
that at the LHC energy the $WW$ fusion constitutes about $10-15 \%$ of 
the integrated inclusive cross section. If the weak boson fusion 
contribution was separated, the measurement of the $WWH$ (or $ZZH$) coupling
would be very interesting test of the Standard Model.
 
In the present paper we are interested in differential distributions
of Higgs boson rather than in the integrated cross section. 

\begin{figure*}
\begin{center}
\includegraphics[width=5cm]{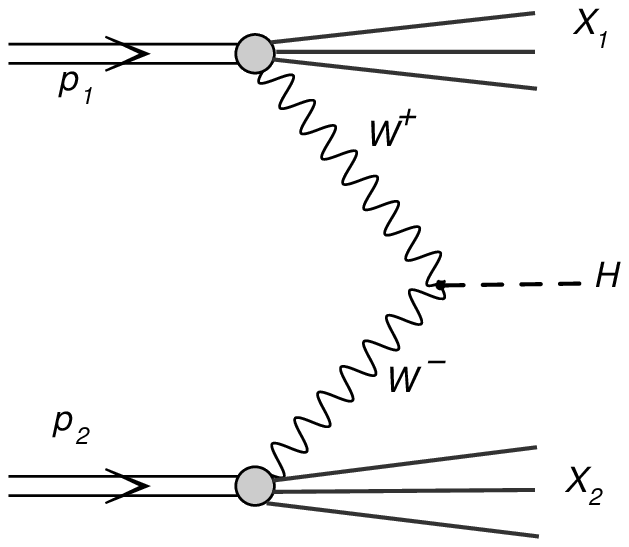}
\includegraphics[width=5cm]{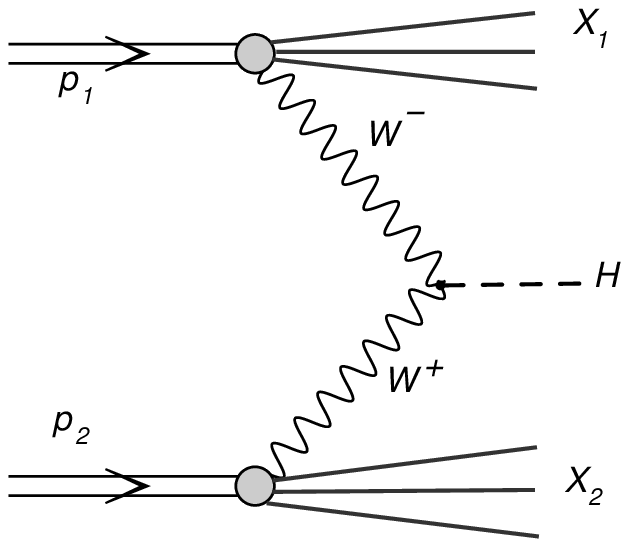}
\caption{Diagrams for the $WW$ fusion.
}
\label{fig:WW_fusion}
\end{center}
\end{figure*}

For the gauge boson fusion the partonic
subprocess is of the 2 $\to$ 3 type:
$q(p_1) + q(p_2) \to q(p_3) + q(p_4) + H(p_H)$
(see Fig.~\ref{fig:WW_fusion}).

The corresponding proton-proton cross section can be written as
\begin{eqnarray}
d \sigma = {\cal F}_{12}^{VV}(x_1,x_2) \; \frac{1}{2 \hat s}
\; \overline{ | {\cal M}_{qq \to qqH} |^2 } \;
\frac{d^3 p_3}{(2 \pi)^3 2 E_3}
\frac{d^3 p_4}{(2 \pi)^3 2 E_4}
\frac{d^3 p_H}{(2 \pi)^3 2 E_H} \nonumber \\
\times \; (2 \pi)^4 \delta^{4}(p_1+p_2-p_3-p_4-p_H) \; d x_1 d x_2 \; .
\label{WW_fusion}
\end{eqnarray}
The next-to-leading order corrections to the matrix element of
the $WW$ fusion are rather small \cite{WW_NLO}.
The leading-order subprocess matrix element was calculated first 
in Ref.~\cite{CD84}.
The spin averaged matrix element squared reads
\begin{equation}
\overline{ | {\cal M} |^2 } = 128 \sqrt{2} G_F^3
\frac{M_W^8 (p_1 \cdot p_2) (p_3 \cdot p_4)}
{(2 p_3 \cdot p_1 + M_W^2)^2 (2 p_4 \cdot p_2 + M_W^2)^2} \; .
\label{M_WW}
\end{equation}
For the WW fusion, limiting to light flavours,
the partonic function is
\begin{eqnarray}
&&{\cal F}_{12}^{WW}(x_1,x_2) = \nonumber \\
&&\left( u_1(x_1,\mu_1^2)+\bar d_1(x_1,\mu_1^2)+\bar s_1(x_1,\mu_1^2) \right)
\left( \bar u_2(x_2,\mu_2^2)+d_2(x_2,\mu_2^2)+s_2(x_2,\mu_2^2) \right) +
\nonumber \\
&&\left( \bar u_1(x_1,\mu_1^2)+d_1(x_1,\mu_1^2)+s_1(x_1,\mu_1^2) \right)
\left( u_2(x_2,\mu_2^2)+\bar d_2(x_2,\mu_2^2)+\bar s_2(x_2,\mu_2^2) \right)
\; .
\label{partonic_function}
\end{eqnarray}
In the following we take $\mu_1^2 = \mu_2^2 = M_H^2$. 
It is convenient to introduce the following new variables:
\begin{equation}
\begin{split}
\vec{p}_{+} = \vec{p}_{3} + \vec{p}_{4} \; , \\
\vec{p}_{-} = \vec{p}_{3} - \vec{p}_{4} \; ,
\end{split}
\label{pplus_pminus}
\end{equation}
which allow to eliminate the momentum-dependent $\delta^3(...)$
in Eq.(\ref{WW_fusion}).
Instead of integrating over $x_1$ and $x_2$ we shall
integrate over
$y_1 \equiv \ln(1/x_1)$ and $y_2 \equiv \ln(1/x_2)$.
Then using Eq.(\ref{WW_fusion}) we can write the inclusive
spectrum of Higgs boson as
\begin{equation}
\begin{split}
\frac{d \sigma}{dy d^2 p_t} =& \int d y_1 d y_2 \;
x_1 x_2 {\cal F}(x_1,x_2,\mu_1^2,\mu_2^2)
\; \frac{1}{2 \hat s} \frac{d^3 p_{-}}{16}
\; \overline{ | {\cal M}_{qq \to qqH} |^2 } \;
\frac{1}{2E_3} \; \frac{1}{2E_4} \; \\
& \times \; \frac{1}{(2 \pi)^5} \; \delta(E_1+E_2-E_3-E_4-E_H) \; .
\end{split}
\label{WW_fusion_red1}
\end{equation}
This is effectively a four-dimensional integral which
can be calculated numerically.

Strong and electroweak corrections to the Higgs boson production
via vector-boson fusion at the LHC were calculated e.g. in 
Ref.~\cite{CDD08} and in Ref.~\cite{BMMZ2010}. 
The corrections are relatively small and in the following analysis
we shall show only leading-order results as a reference to
the $k_t$-factorization result.

\subsection{$\bm{ZZ}$ fusion}

\begin{figure*}
\begin{center}
\includegraphics[width=5cm]{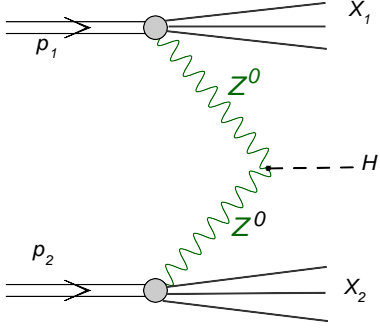}
\caption{Typical diagram for the $ZZ$ fusion.
}
\label{fig:ZZ_fusion}
\end{center}
\end{figure*}

The $ZZ$ fusion (see Fig.~\ref{fig:ZZ_fusion}) can be calculated
in an analogous way. 

The corresponding matrix element depends on the
subprocess type (set of quark, antiquark flavours). 
It can be written as \cite{CD84}:
\begin{equation}
\overline{ | {\cal M} |_{f_1 f_2}^2 } = 128 \sqrt{2} G_F^3 M_Z^8
\frac{C_1^Z(f_1 f_2) (p_1 \cdot p_2) (p_3 \cdot p_4) 
    + C_2^Z(f_1 f_2) (p_1 \cdot p_4) (p_2 \cdot p_3)}
{(2 p_3 \cdot p_1 + M_Z^2)^2 (2 p_4 \cdot p_2 + M_Z^2)^2} 
\; .
\label{M_ZZ}
\end{equation}
The flavour dependent coefficients read:
\begin{eqnarray}
C_1^Z (f_1 f_2) &=& \frac{1}{4} 
\left(  (V_{f_1} - A_{f_1})^2 (V_{f_2} - A_{f_2})^2
      + (V_{f_1} + A_{f_1})^2 (V_{f_2} + A_{f_2})^2 \right) 
\; , \nonumber \\
C_2^Z (f_1 f_2) &=& \frac{1}{4}
\left(  (V_{f_1} - A_{f_1})^2 (V_{f_2} + A_{f_2})^2
      + (V_{f_1} + A_{f_2})^2 (V_{f_2} - A_{f_2})^2 \right) 
\; .
\label{C1Z_C2Z_coefficients}
\end{eqnarray}
The $V_f$ and $A_f$ are well known vector and axial-vector couplings
of the $Z^0$ boson to quarks/antiquarks. They can be expressed in terms 
of third component of the weak isospin, charge of quark/antiquark and
sinus of the Weinberg angle.

The differential cross section is calculated in exactly the same way 
as for the $WW$ fusion.

\subsection{Associated production with $W$ and 
$Z$ bosons}

For completeness one could include also production of the Higgs boson
associated with gauge bosons $W^+$, $W^-$ and $Z^0$.
These are formally lower-order (2 $\to$ 2) processes than 
the $WW$ and $ZZ$ fusion (2 $\to$ 3) processes considered above. 
They were first considered in Ref.~\cite{GNY78}.

The matrix elements are very simple:
\begin{eqnarray}
|{\cal M}_{f_1 f_2 \to W H}|^2 &=& 
\frac{(G_F M_W^2)^2}{72 \pi^2} |V_{f_1 f_2}|^2 
\frac{3 M_W^2 + p_W^2}{({\hat s}-M_W^2)^2} \; , \nonumber \\
|{\cal M}_{f f \to Z H}|^2 &=& \frac{(G_F M_Z^2)^2}{72 \pi^2} 
(V_f^2 + A_f^2) \frac{3 M_Z^2 + p_Z^2}{({\hat s}-M_Z^2)^2} \; .
\label{ME2_associated_production}
\end{eqnarray}
In the equation above $p_V$ is momentum of the gauge boson in 
the  $HV$ center of mass frame:
\begin{equation}
p_W^2 = \frac{1}{4 \hat s}
\left(
{\hat s}^2 + M_V^4 + M_H^4 - 2 {\hat s} M_V^2 - 2 {\hat s}M_H^2 
- 2 M_V^2 M_H^2
\right) \; ;
\label{p_V2}
\end{equation}
where $V = W, Z$.

The fully differential cross section can be written as:
\begin{eqnarray}
\frac{d \sigma}{d y_H d y_W d^2 p_t} &=&  \frac{1}{16 \pi {\hat s}^2} \; |{\cal M}_{f_1 f_2 \to W H}|^2  \nonumber \\
&& \times \; \sum\nolimits_{f_1 f_2} \left(
x_1 q_{f_1}(x_1,\mu^2) x_2 {\bar q}_{f_2}(x_2,\mu^2) +
x_1 \bar q_{f_1}(x_1,\mu^2) x_2 q_{f_2}(x_2,\mu^2)
\right) \; , \nonumber \\
\frac{d \sigma}{d y_H d y_Z d^2 p_t} &=& \frac{1}{16 \pi {\hat s}^2} \; |{\cal M}_{f f \to Z H}|^2 \nonumber \\
&& \times \sum\nolimits_{f} \left(
x_1 q_{f}(x_1,\mu^2) x_2 {\bar q}_{f}(x_2,\mu^2) +
x_1 \bar q_{f}(x_1,\mu^2) x_2 q_{f}(x_2,\mu^2)
\right) \; .
\label{qqbar_VH}
\end{eqnarray}
The Higgs boson distributions can be obtained from those above
by integrating over $y_W$ and $y_Z$, respectively. 

\section{Results}

\subsection{$\bm{g g \to H}$ and subsequent $\bm{H \to \gamma \gamma}$ 
decay}
\begin{table}
\caption{The cross section for Higgs production $p_t <$ 400 GeV in pb 
for $\sqrt{s}$ = 8 TeV and for different UGDFs from the literature.
For comparison we show also contribution of the $g g \to g H g$ and $i j \to i H j$
processes ($p_{1t},p_{2t} >$ 10 GeV), and $WW$ and $ZZ$ fusion.}
\label{table:Higgs_total_cross_section}
\begin{tabular}{|c|c|}
\hline
contribution               & $\mu_r^2=\mu_f^2=m_H^2$ \\
\hline 
KMR                       &  5.2349  \\   
Jung CCFM (set$A0$)            &  8.2705  \\ 
Jung CCFM (set$A+$)           & 12.3791  \\
Jung CCFM (set$A-$)           &  5.7335  \\            
Kutak-Sta{\'s}to            &  2.6074  \\
Kutak-Sapeta              &  1.5465  \\
\hline
KMR, $q_{1t}$, $q_{2t} >$ 10 GeV               &  2.4585  \\     
\hline
$g g \to g H g$,  $q_{1t},q_{2t} >$ 10 GeV  & 0.24   \\
$i j \to i H j$,  $q_{1t},q_{2t} >$ 10 GeV  & 0.57   \\
\hline
$WW$ fusion                 & 0.9332  \\
$ZZ$ fusion                 & 0.02641 \\
\hline
\end{tabular}
\end{table}

In Table \ref{table:Higgs_total_cross_section} we present
total (integrated over full phase space) cross section 
for the $2 \to 1$ gluon-gluon fusion mechanism for several UGDFs from the
literature at $\sqrt{s}$ = 8 TeV. For reference the leading-order
collinear approximation result is typically 5-7 pb depending somewhat 
on parton distribution functions used in the calculation.
The $k_t$-factorization results (for several UGDFs used here) 
are somewhat smaller. There are two reasons for this. 
First, when calculating gluon longitudinal momentum
fractions transverse momentum of the Higgs boson is included 
which increases $x_1$ and $x_2$ and therefore lowers the cross section.
Secondly, many low-$x$ UGDF do not apply and/or are too small 
in the region of $x_1, x_2 >$ 0.01.
Quite different cross sections are obtained for different UGDFs.
This shows that the UGDFs (often fitted only to HERA data) are much 
more uncertain than the collinear gluon distribution functions (GDF) 
fitted to many sets of high-energy data. However, UGDFs have advantage 
that they can be used for correct (exclusive) kinematics including 
transverse momenta of initial gluons, which cannot be addressed 
properly in collinear calculations.

For comparison in the middle block we show contribution 
of $i j \to i H j$ processes calculated in the collinear-factorization
approach for the jet transverse momenta bigger than 10 GeV. 
The $g g \to g H g$ contribution is of similar size as
that for the leading order $g g \to H$ $k_t$-factorization approach. We think that the latter
contribution is to large extent contained in the calculation with the KMR UGDF.
However, the quark and antiquark initiated contributions which are
also fairly large ($\sim$ 0.6 pb) must be included in addition explicitly.

At the very bottom we show contributions of the $WW$ and $ZZ$ fusions.
The electroweak contribution is quite sizeable. As will be shown
below they play important role at large Higgs boson transverse momenta.

\begin{figure}[!h]
\includegraphics[width=7cm]{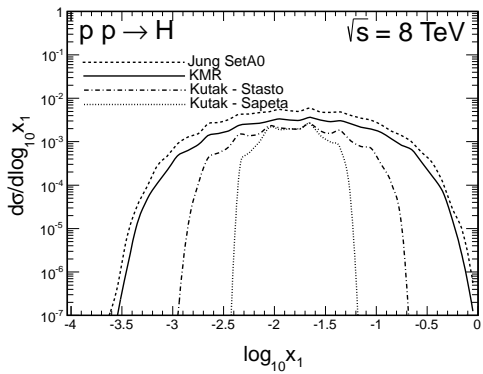}
   \caption{
\small Distribution in $\log_{10}(x_1)$ or $\log_{10}(x_2)$ for $gg\to H$ and
for different UGDFs used in the present analysis.
}
 \label{fig:log10x}
\end{figure}

The so different cross sections obtained with different UGDFs 
may be partially understood by looking at distribution 
in $x_1$ or $x_2$ (see Fig.~\ref{fig:log10x}).
The KMR UGDF gives much larger contribution in the region of 
$x_1, x_2 >$ 0.01 than the typically small-$x$ UGDFs. The other UGDFs are in this range of $x$'s not
very realistic.

\begin{figure}[!h]
\includegraphics[width=5.5cm]{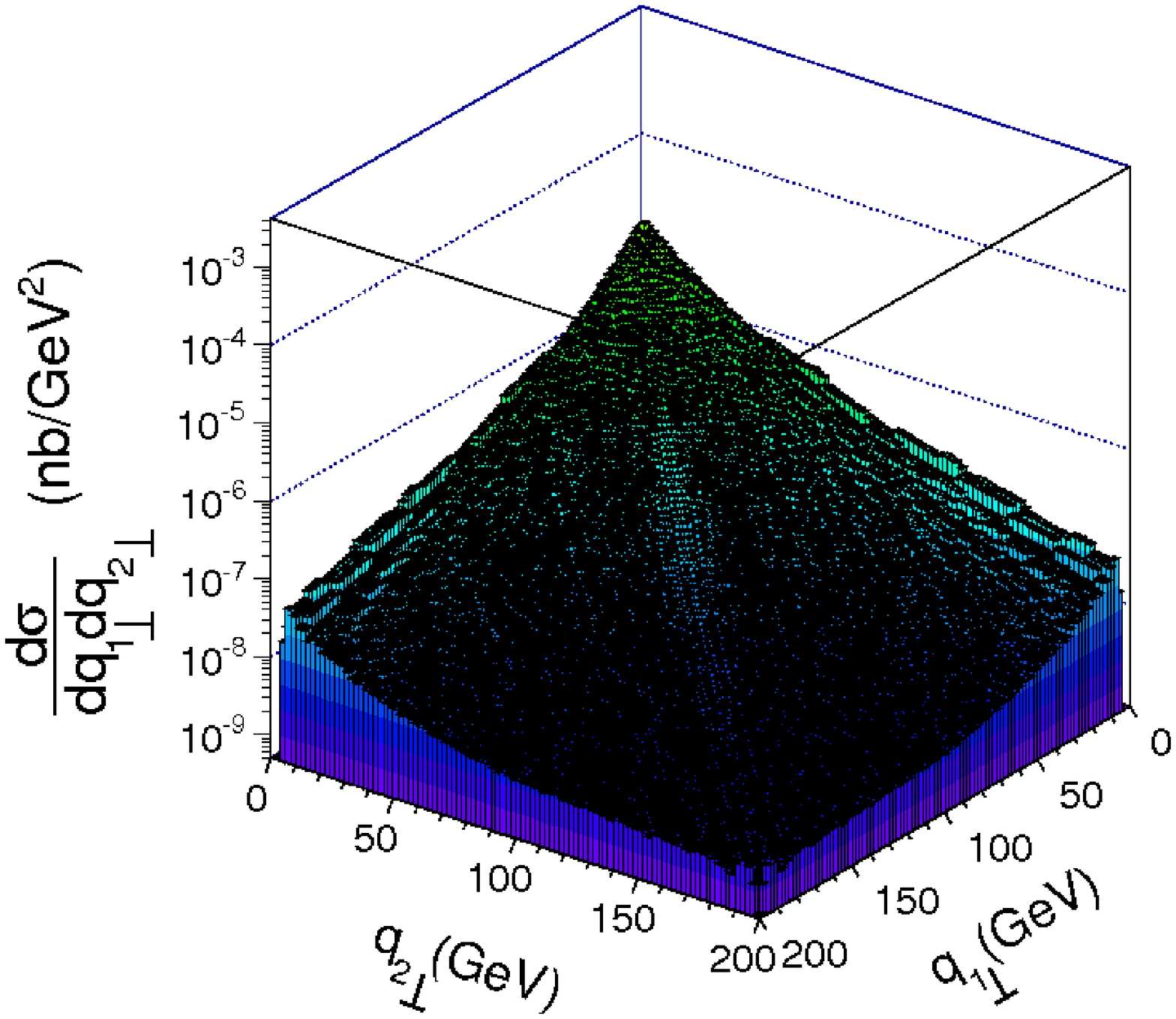}
\includegraphics[width=5.5cm]{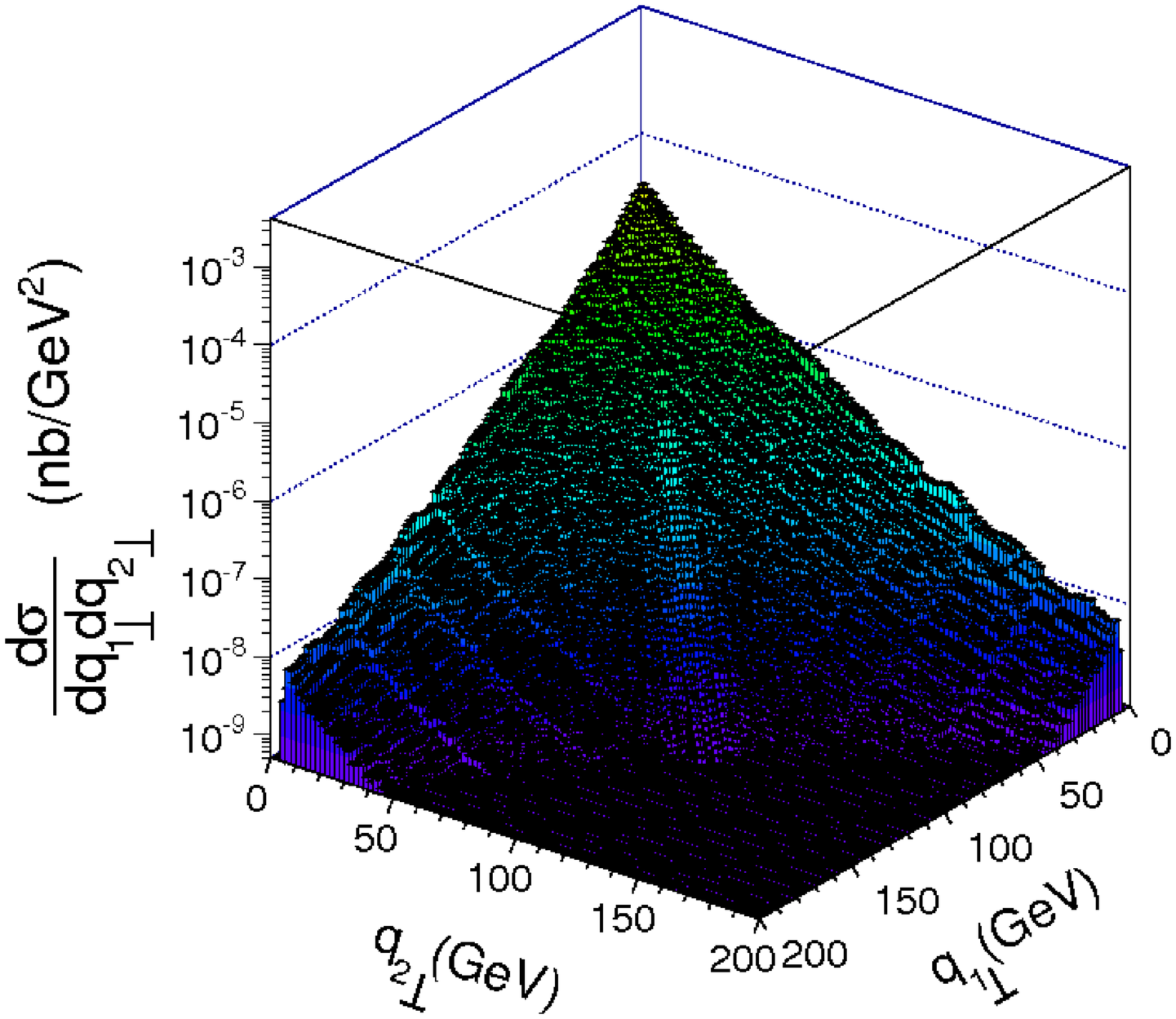}
\includegraphics[width=5.5cm]{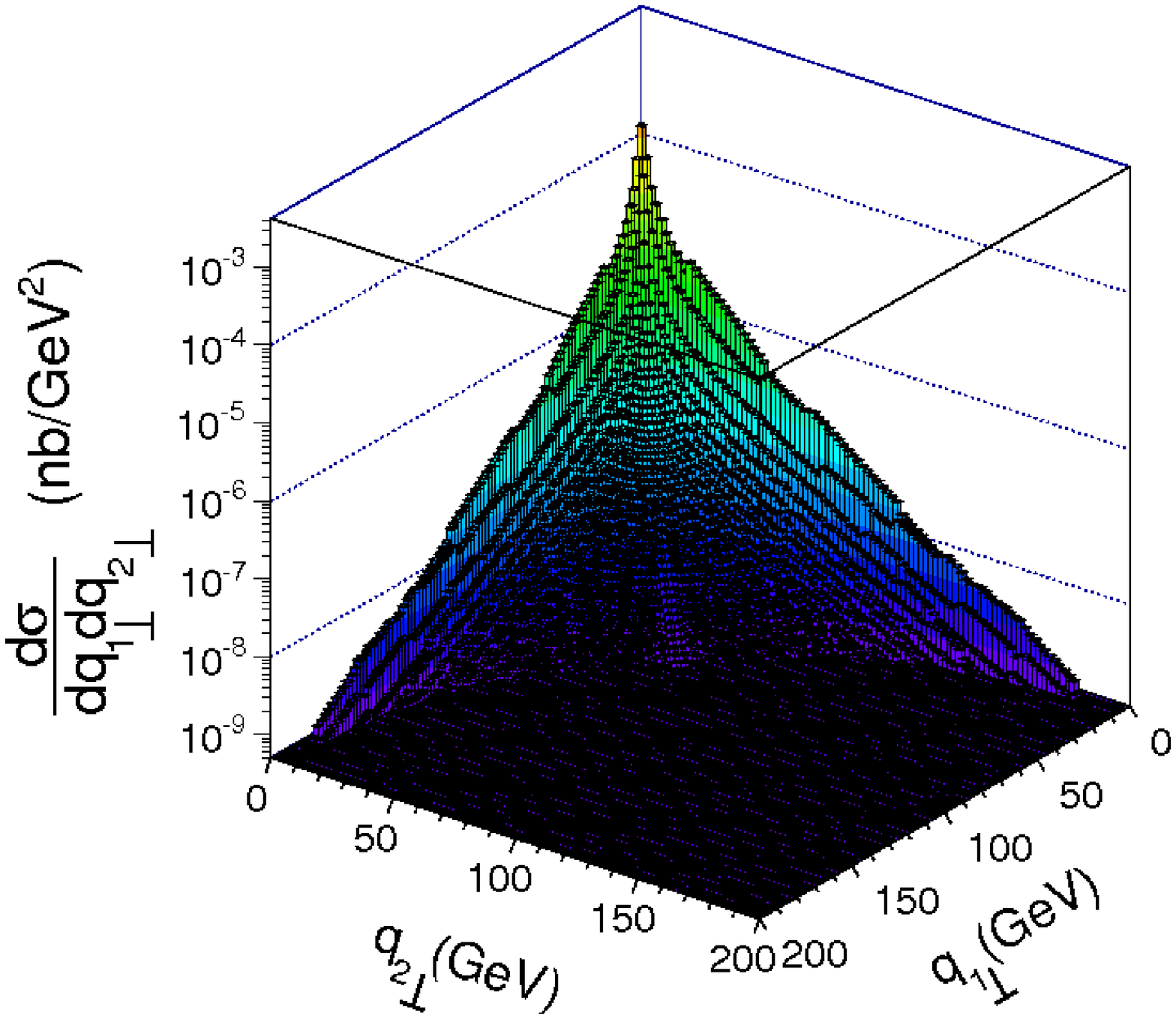}
\includegraphics[width=5.5cm]{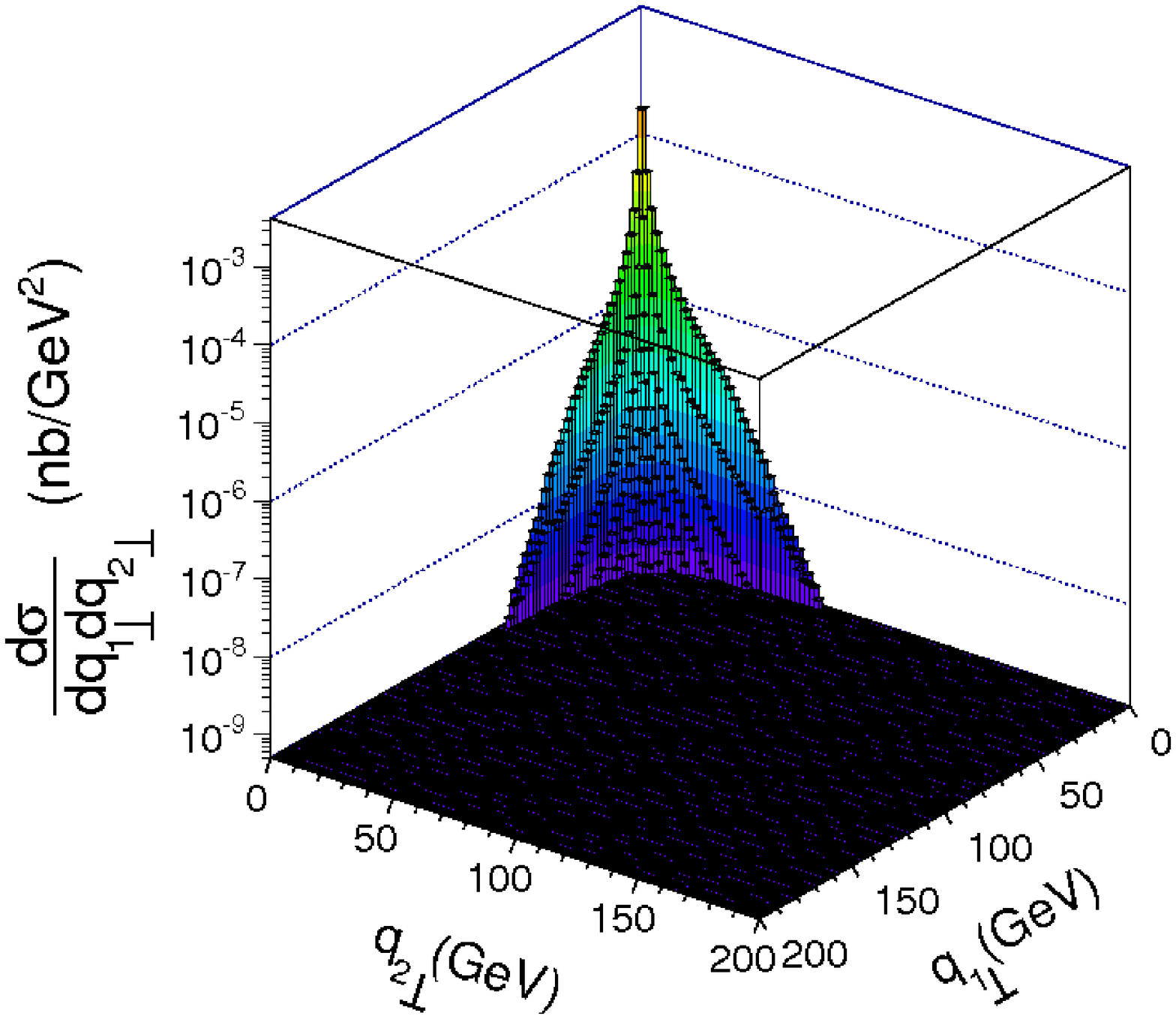}
   \caption{
\small Distribution in $q_{1t}$ and $q_{2t}$ for $gg \to H$ and for different
UGDFs: KMR, Jung CCFM (set$A0$), Kutak-Sta\'sto
and Kutak-Sapeta.
}
 \label{fig:q1tq2t}
\end{figure}

In addition, the different UGDFs in the literature has quite 
different dependence on gluon transverse momenta. This is well demonstrated in 
Fig.~\ref{fig:q1tq2t} where we show two-dimensional maps in 
$q_{1t} \times q_{2t}$ for different UGDFs. The Kutak-Sapeta UGDF
gives a sharp peak at large $q_{1t}$ and $q_{2t}$. This means that
using such an UGDF one cannot obtain large Higgs boson transverse momenta.
Quite large gluon transverse momenta ($q_{1t},q_{2t} \sim m_H$) enter 
the production of the Higgs boson for the KMR and Jung CCFM (set$A0$) UGDFs. 
For the KMR UGDF a clear enhancement at small $q_{1t}$ or $q_{2t}$
can be observed. 
This is rather a region of nonperturbative nature, where the KMR UGDF 
is rather extrapolated than calculated.
However, we have checked that the contribution of the region
when $q_{1t} <$ 2 GeV or $q_{2t} <$ 2 GeV constitutes only less than $5 \%$
of the integrated cross section. This is then a simple estimate
of uncertainty of the whole approach.

\begin{figure}[!h]
\includegraphics[width=5.5cm]{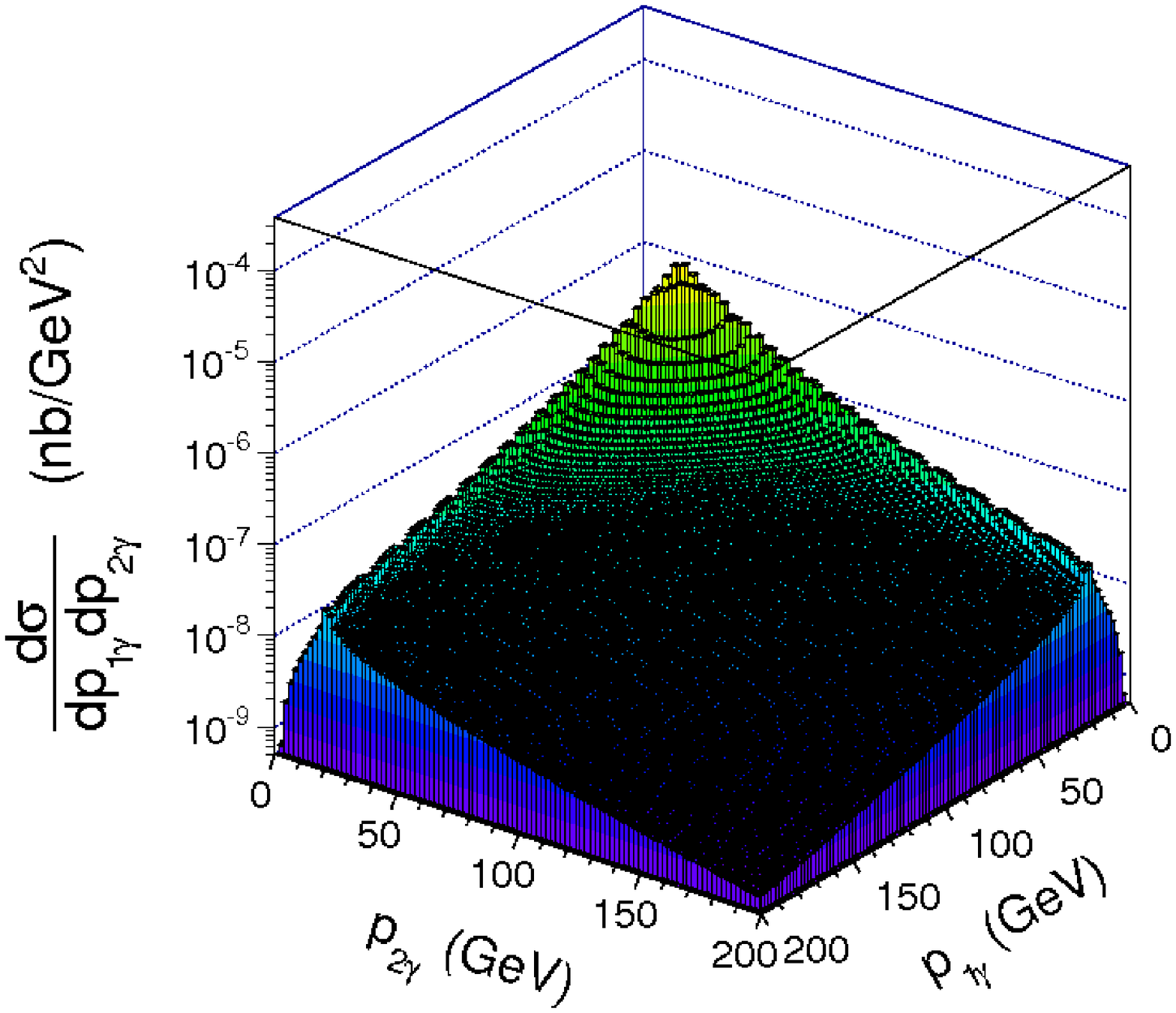}
\includegraphics[width=5.5cm]{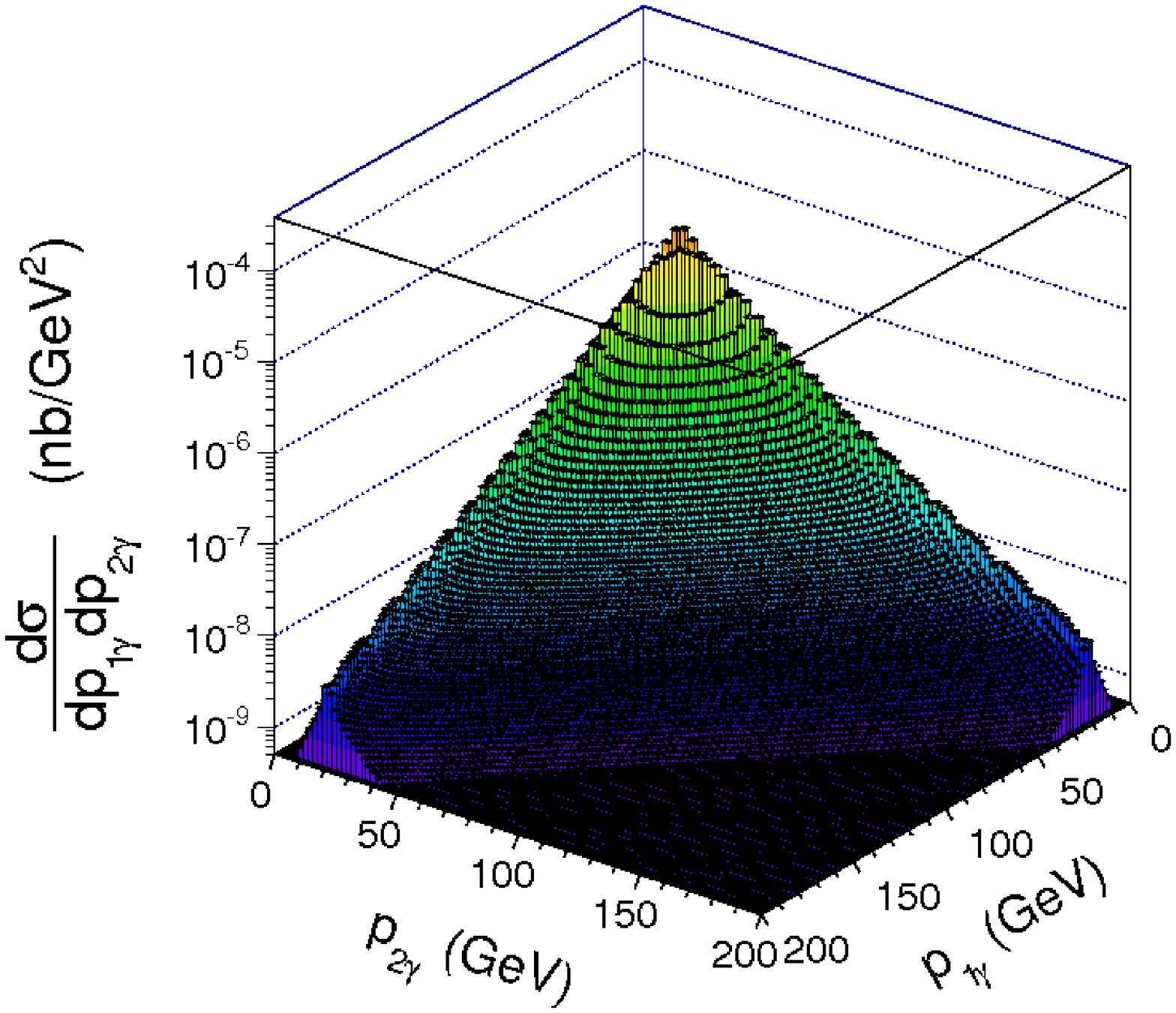}\\
\includegraphics[width=5.5cm]{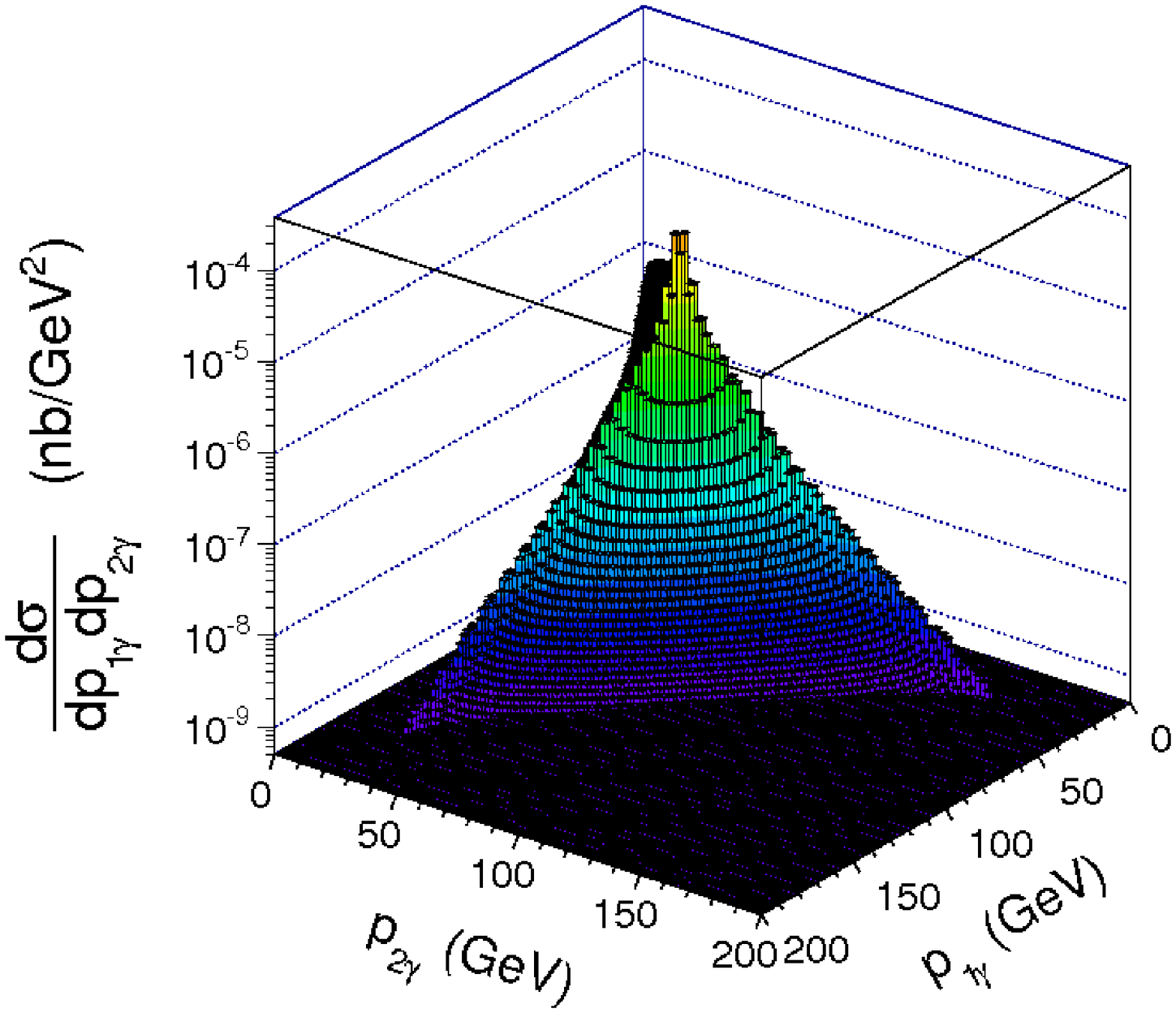}
\includegraphics[width=5.5cm]{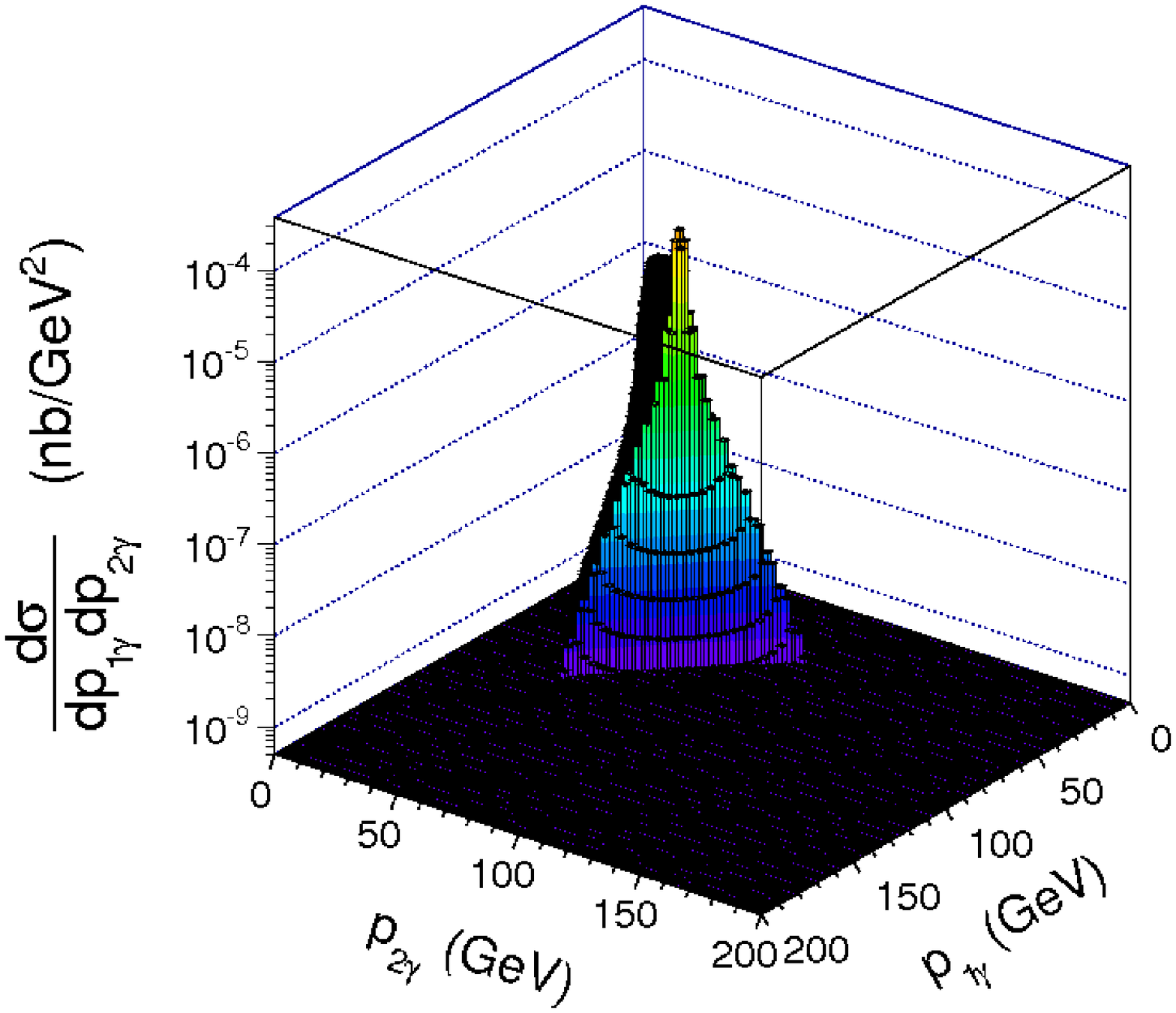}\\

   \caption{
\small Distributions in photon transverse momenta $p_{1t}$ and $p_{2t}$ for the $gg\to H$ and 
for the KMR, Jung CCFM (set$A0$),
Kutak-Sta{\'s}to and Kutak-Sapeta UGDFs.
}
 \label{fig:p1tp2t}
\end{figure}

\begin{figure}[!h]
\includegraphics[width=6cm]{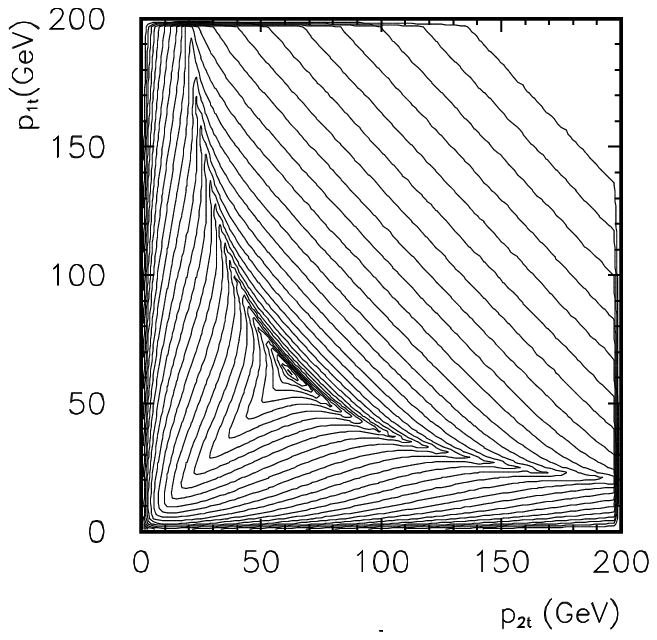}
\includegraphics[width=6cm]{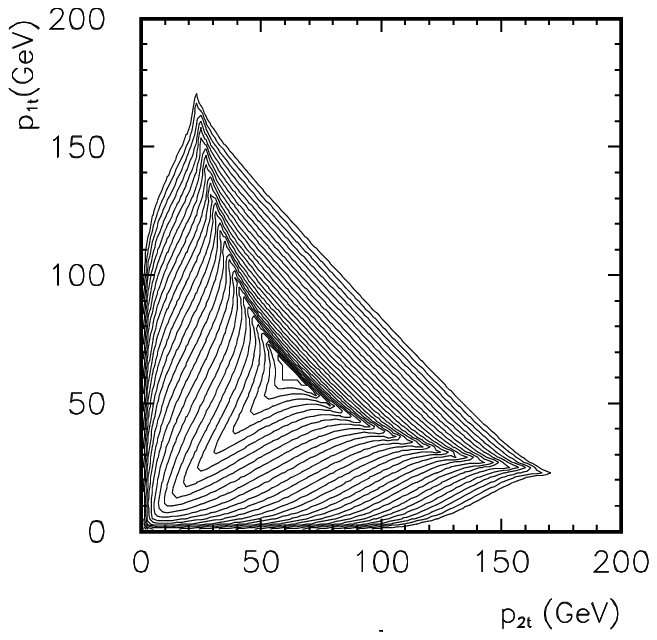}
   \caption{
\small Distributions in photon transverse momenta $p_{1t}$ and $p_{2t}$ for $gg \to H$ and
for the KMR and Jung CCFM (set$A0$) UGDFs for the contour representation.
}
 \label{fig:p1tp2t_map}
\end{figure}

Now we can proceed to the production of photons. We start from two-dimensional
distributions in $\frac{d \sigma}{d y d p_t}$ in rapidity and transverse
momentum of the Higgs boson calculated according to Eq.(\ref{useful_formula}) 
and perform its decay isotropically in the Higgs boson rest frame (assuming spin zero of the Higgs boson). 
Next relativistic boosts are performed to get distributions of photons 
in the proton-proton center of mass system.
As an example in Fig.~\ref{fig:p1tp2t} we show two-dimensional 
distributions in photon transverse momenta. Also here the distributions
for different UGDFs differ significantly.
%
%
In Fig.~\ref{fig:p1tp2t_map} we show in addition two examples
but in the contour form which shows some details better than the lego plot.

\begin{figure}[!h]
\includegraphics[width=8cm]{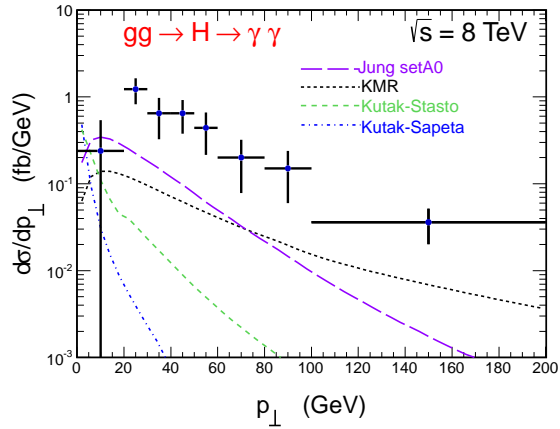}
   \caption{
\small Transverse momentum distribution of the Higgs boson
produced in the $g g \to H$ subprocess
in the $\gamma \gamma$ channels for different UGDFs from
the literature.
}
 \label{fig:pt_gamgam_UGDFs}
\end{figure}

In order to confront our calculations with the preliminary ATLAS data
\cite{ATLAS_Higgs} extra cuts on photon rapidities and transverse
momenta must be imposed in addition. We require:
\begin{equation}
-2.37 <  \eta_{\gamma,1}, \eta_{\gamma,2} < 2.37 \nonumber ,
\end{equation}
\begin{equation}
\max \left( p_{1t},p_{2t} \right) > 0.35 \times M_{\gamma \gamma},
 \;\;\; \min \left( p_{1t},p_{2t} \right) > 0.25 \times M_{\gamma \gamma}\nonumber ,
\end{equation}
\begin{equation}
105 \; \textrm{GeV} < M_{\gamma \gamma} < 160 \; \textrm{GeV}
\end{equation}
as relevant for the ATLAS analysis \cite{ATLAS_Higgs}.
The distribution in transverse momentum of the photon pair 
(almost transverse momentum of the Higgs boson)
is shown in Fig.~\ref{fig:pt_gamgam_UGDFs} for different UGDFs from 
the literature together with the ATLAS data \cite{ATLAS_Higgs}.
The calculated distributions lay much below the ATLAS data
in clear disagreement with the recent calculation in Ref.~\cite{LMZ2014}.
We shall return to the discussion of the disagreement and its potential
explanation in the rest part of the paper.

\subsection{$g g \to H^* \to \gamma \gamma$}

In this section we shall present results of calculations performed
within the $k_t$-factorization in the second method. The photon
distributions from virtual Higgs decay are calculated including
correctly kinematics of the $2 \to 2$ subprocess 
$g g \to H^* \to \gamma \gamma$. Now we wish to compare differential cross sections obtained in this way with those
obtained within the first method. Clearly the second method leads to 
sizeably larger cross sections. This may be helpful in the
context of the deficit discussed in the previous section, but certainly not sufficient.

\begin{figure}[!h]
\includegraphics[width=8cm]{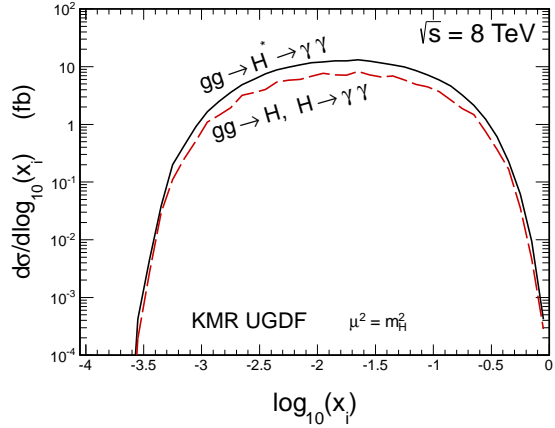}
   \caption{
\small Distribution in $\log_{10} x_i$ for the KMR UGDF
and $\mu^2 = m_{H}^2$ for first (on-shell Higgs boson, long-dashed line) 
and second (off-shell Higgs boson, solid line) method.
}
\label{fig:dsig_dlog10x}
\end{figure}

In Fig.~\ref{fig:dsig_dlog10x}, as an example, we show somewhat theoretical
distribution in $\log_{10}(x_i)$, $i = 1, 2$ for 
the KMR UGDF with $\mu^2 = m_H^2$.
Both low-$x$ ($x < 10^{-2}$) and high-$x$ ($x > 10^{-2}$) regions give similar contributions to the cross section.

\begin{figure}[!h]
\includegraphics[width=8cm]{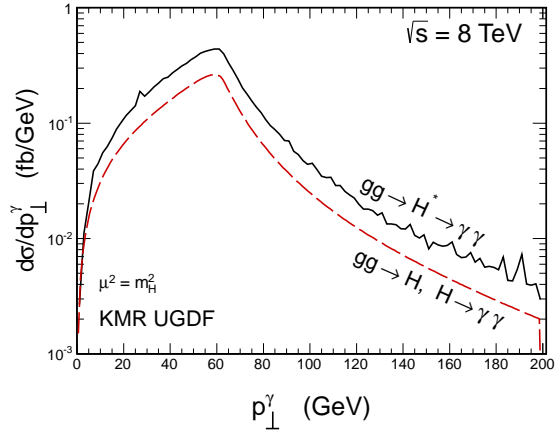}
   \caption{
\small Distribution in $p_{it}$ (i=1,2) for the KMR UGDF
and $\mu^2 = m_{H}^2$ for the first (long-dashed line) and second (solid
line) method.
}
 \label{fig:dsig_dpit}
\end{figure}

In Fig.~\ref{fig:dsig_dpit} we show distribution in 
$p_{1t}$ or $p_{2t}$ (identical) for the two methods. The two
distributions are rather similar as far as the shape is considered.

\begin{figure}[!h]
\includegraphics[width=8cm]{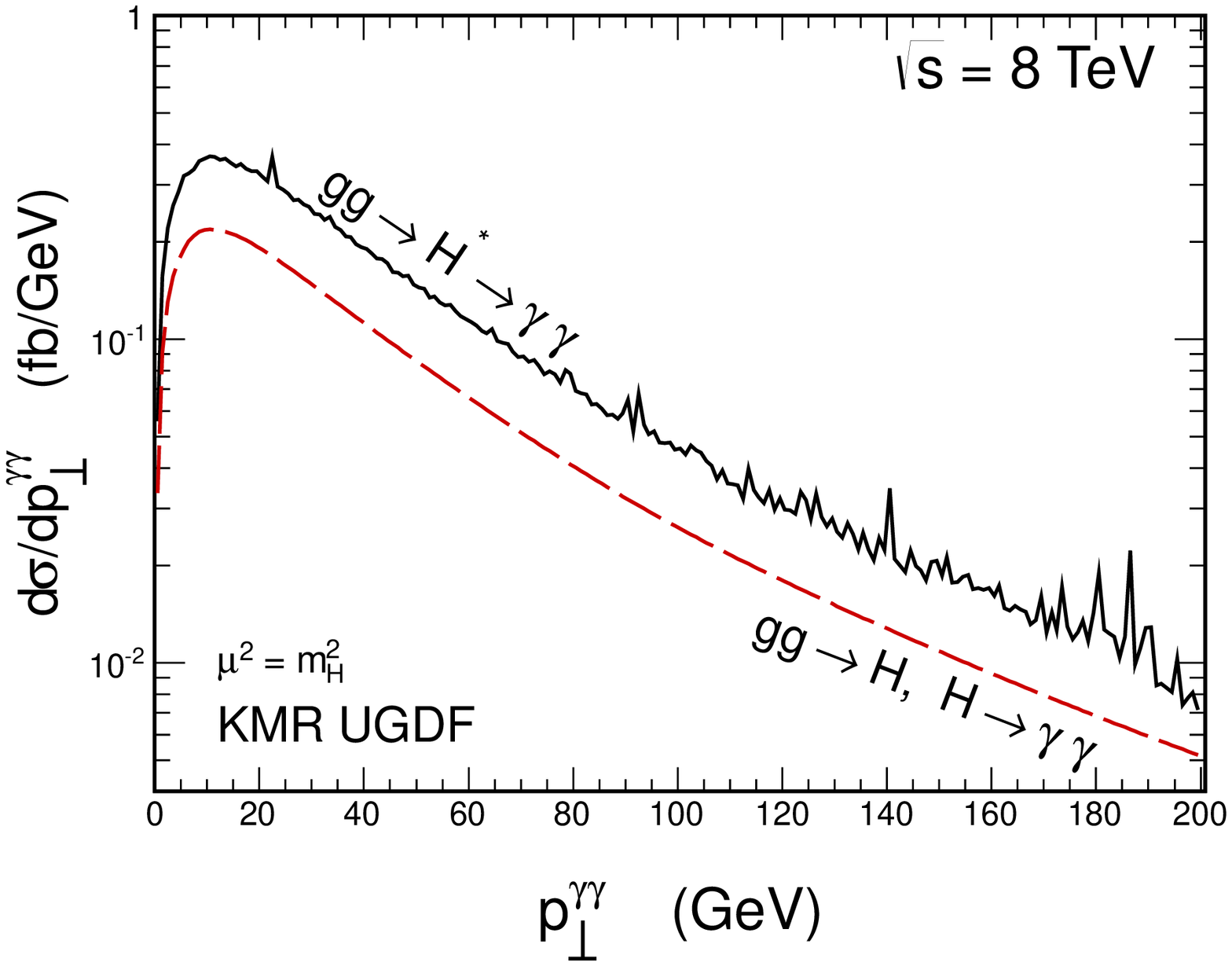}
\caption{
\small Distribution in $p_{t,sum}$ for the KMR
UGDF and $\mu^2 = m_{H}^2$ for the first (long-dashed line) and second 
(solid line) method.
}
\label{fig:dsig_dptsum}
\end{figure}

The distribution in $p_{t,sum}$ ($\vec{p}_{t,sum} = \vec{p}_{1t} + \vec{p}_{2t}$) is particularly interesting as it reflects distribution of the Higgs boson and can 
be measured experimentally. In Fig.~\ref{fig:dsig_dptsum} we again
compare results obtained in the two methods. The shapes obtained
with the two methods are practically identical but there is a small
difference in the normalization.

\begin{figure}[!h]
\includegraphics[width=8cm]{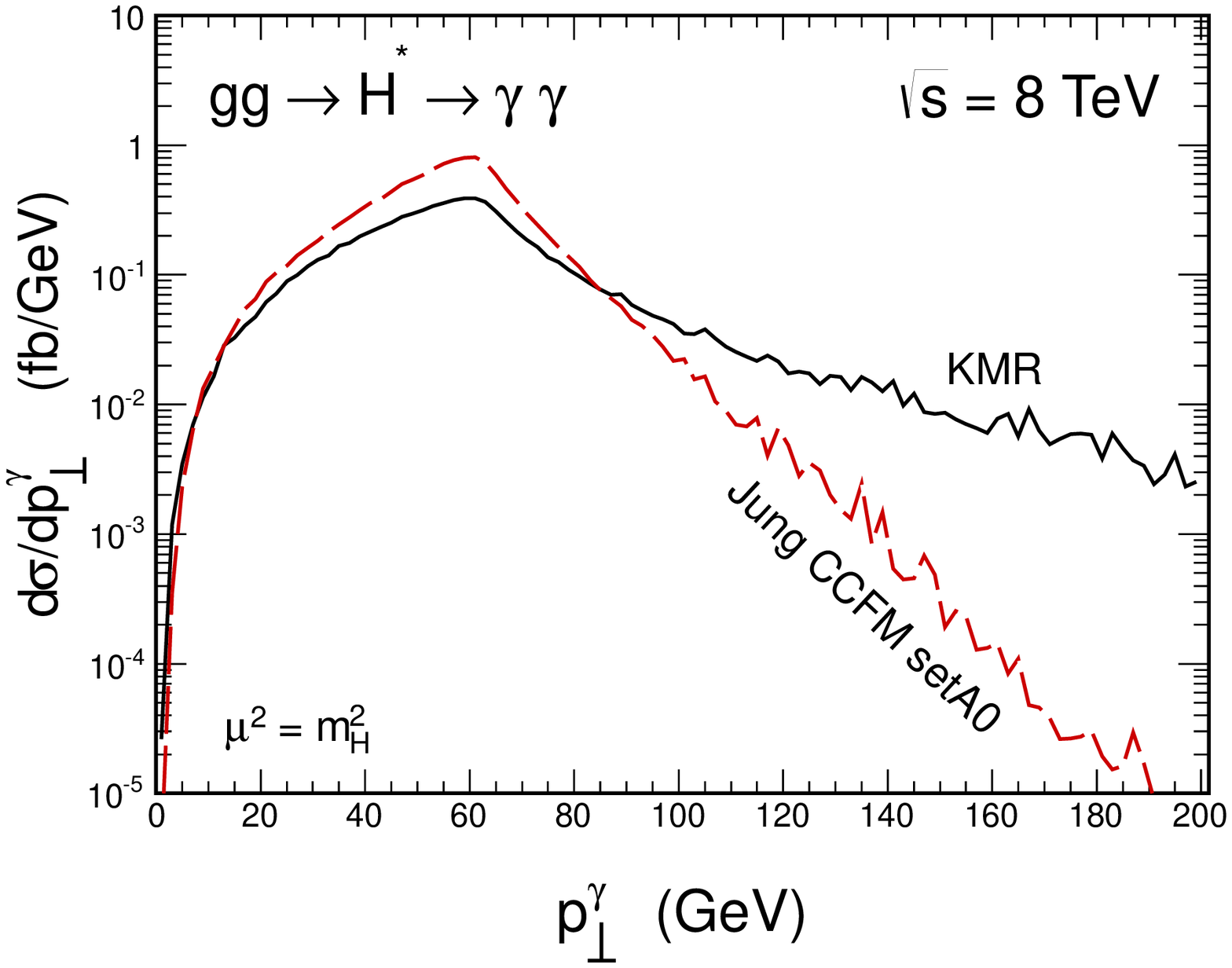}
\caption{
\small Distribution of photon transverse momentum for the 
KMR (solid line) and the Jung CCFM (set$A0$) (long-dashed line) UGDF, for $\mu^2 = m_{H}^2$.
}
\label{fig:dsig_dpt1}
\end{figure}

Now we wish to show several results for the second approach only.
Let us start from single photon transverse momentum distribution.
In Fig.~\ref{fig:dsig_dpt1} we show such distributions for
two selected UGDFs. The peak at $p_t \sim m_H/2$ is of kinematical
nature. The KMR UGDF leads to larger photon transverse momenta.

\begin{figure}[!h]
\includegraphics[width=8cm]{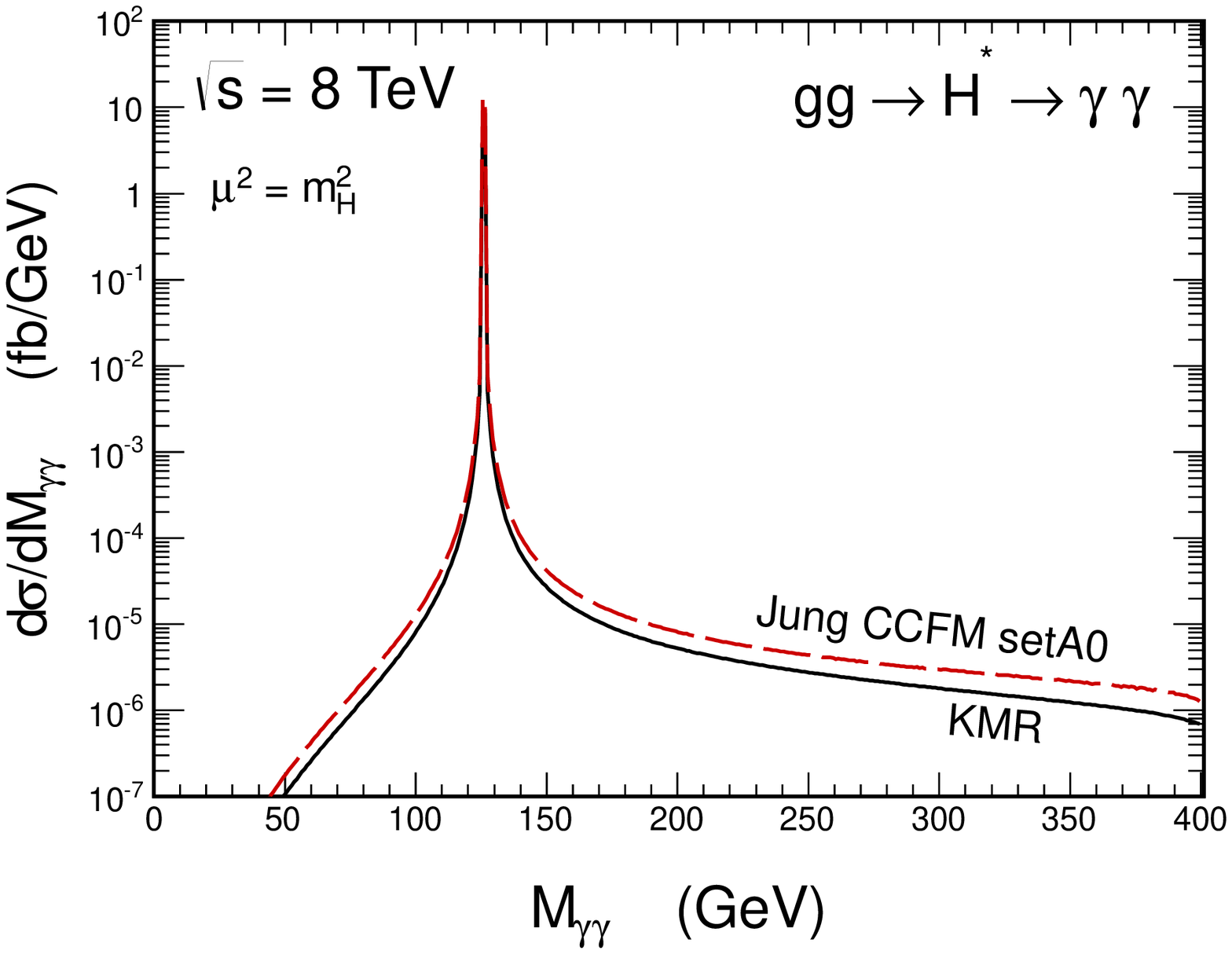}
\caption{
\small Distribution of diphoton invariant mass for the 
KMR (solid line) and the Jung CCFM (set$A0$) (long-dashed line) UGDF, 
for $\mu^2 = m_{H}^2$.
}
\label{fig:dsig_dMgamgam}
\end{figure}

Particularly interesting is distribution in two-photon invariant mass.
The huge peak at $M_{\gamma \gamma} = M_H$ corresponds to on-shell
Higgs boson. We observe (see Fig.~\ref{fig:dsig_dMgamgam}) small 
contributions from off-shell Higgs boson configurations with invariant masses 
both smaller or larger than the on shell (peak) value. 
The sharp peak shows that the integration of the cross section
is not easy. We have, however, carefully checked the convergence.

\begin{figure}[!h]
\includegraphics[width=8cm]{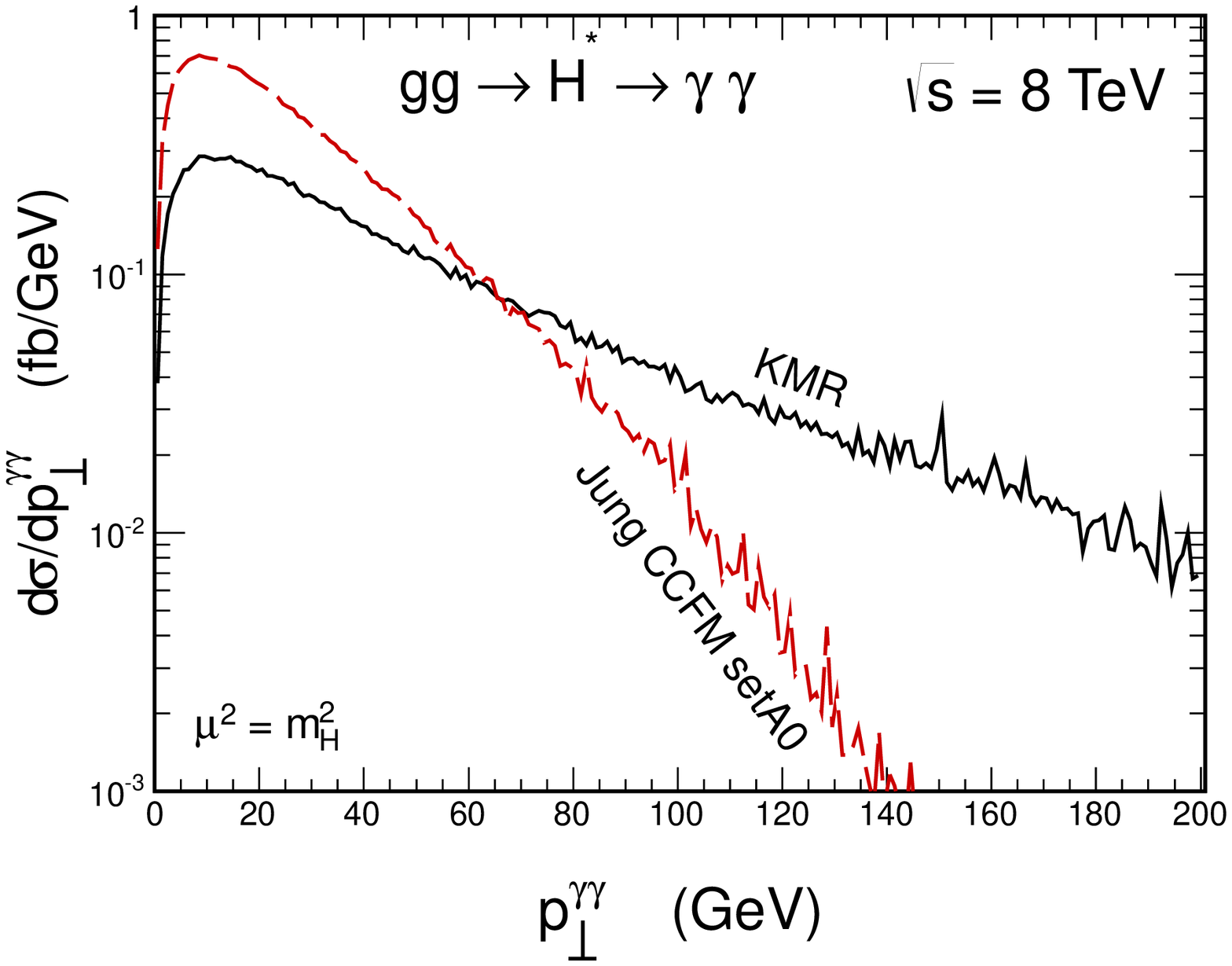}
\caption{
\small Distribution of diphoton transverse momentum for the 
KMR (solid line) and the Jung CCFM (set$A0$) (long-dashed line) UGDF, for $\mu^2 = m_{H}^2$.
}
\label{fig:dsig_dptsum_ugdf}
\end{figure}

As was already mentioned, the distribution for $p_{t,sum}$ reflects the 
Higgs boson transverse momenta. Interesting question is how the
distribution is sensitive to the choice of the UGDF model. 
Figure~\ref{fig:dsig_dptsum_ugdf} shows that the KMR UGDF generates 
much bigger Higgs boson transverse momenta than the Jung CCFM (set$A0$).

\begin{figure}[!h]
\includegraphics[width=8cm]{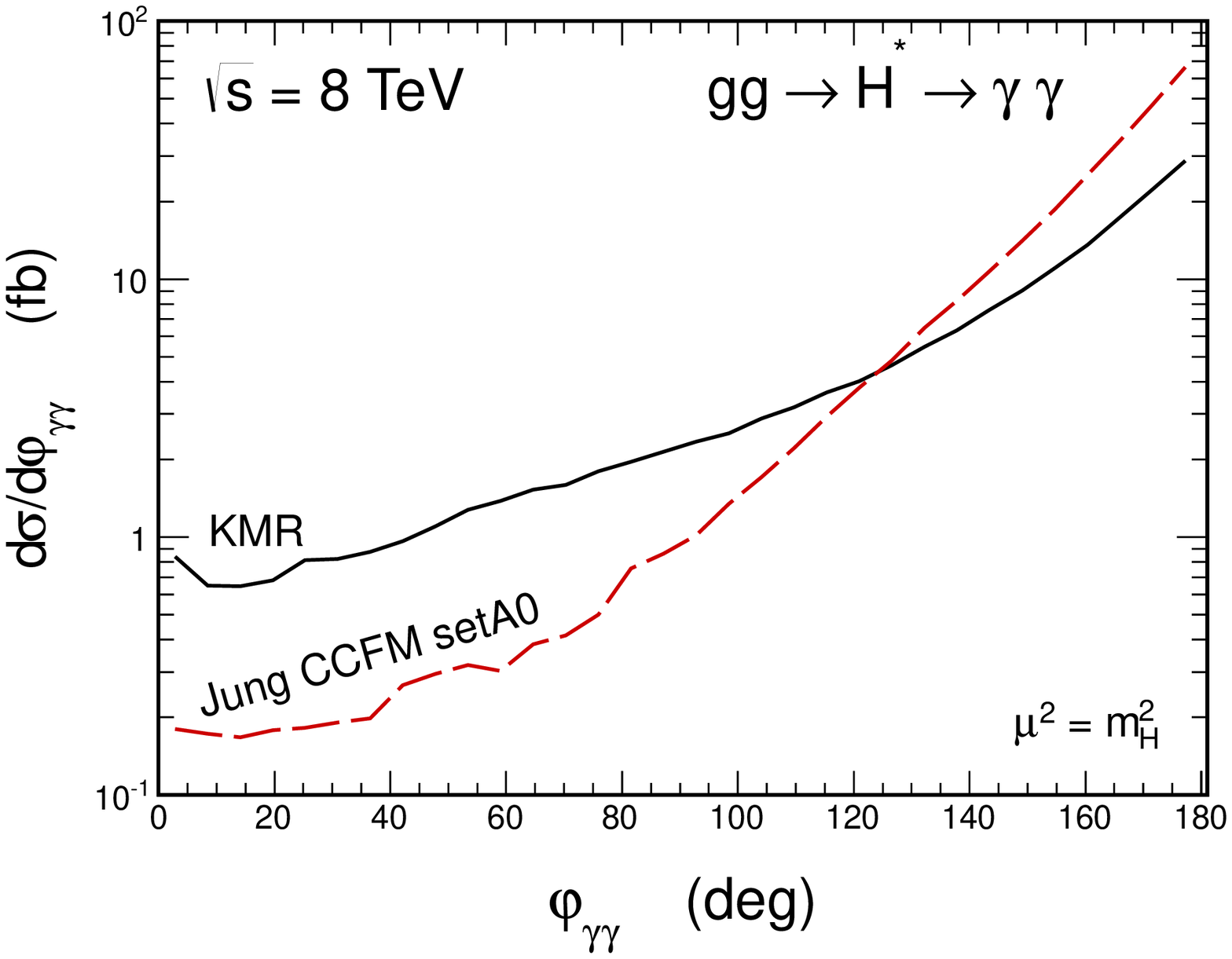}
\caption{
\small Distribution of azimuthal angle between photons for the 
KMR (solid line) and the Jung CCFM (set$A0$) (long-dashed line) UGDF, 
for $\mu^2 = m_{H}^2$.
}
\label{fig:dsig_dphigamgam}
\end{figure}

Another interesting observable is correlation in azimuthal angle 
between the outgoing photons (see Fig.~\ref{fig:dsig_dphigamgam}). 
A bigger back-to-back correlation is observed for the Jung CCFM (set$A0$)
than for the KMR UGDF. This is similar as already observed
for azimuthal correlations between $c \bar c$ (see e.g. Ref.~\cite{MS2013}).
The decorrelation for the KMR UGDF is even larger (compare only shapes) 
than in the soft-gluon transverse momentum resummation \cite{FFGT2012}.
Small $\phi_{\gamma \gamma}$ are strongly correlated with large
gluon transverse momenta $q_{1t}$ or $q_{2t}$. As discussed above this
may be overestimated in the $k_t$-factorization approach with the KMR UGDF.

\begin{figure}[!h]
\includegraphics[width=6cm]{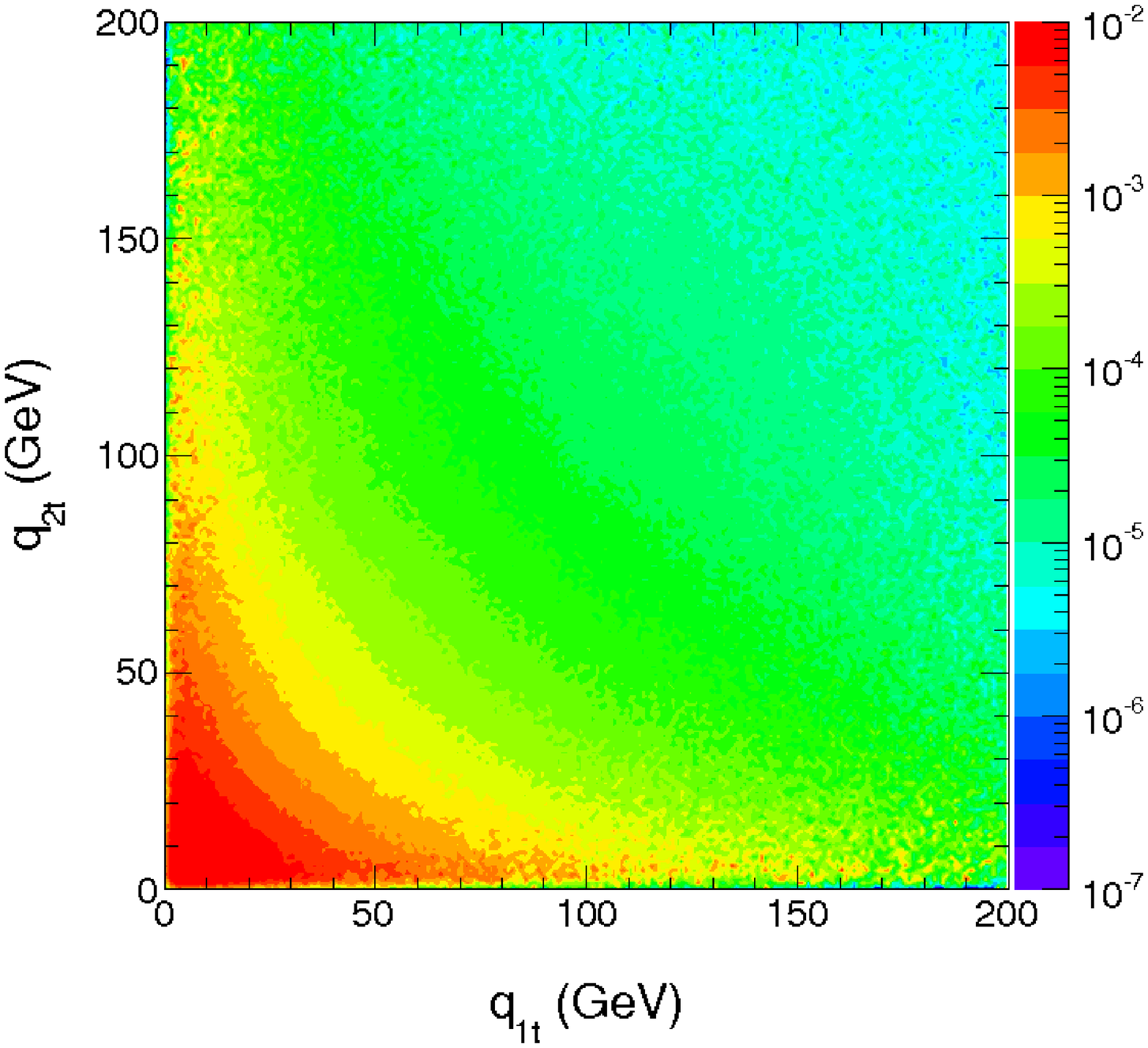}
\includegraphics[width=6cm]{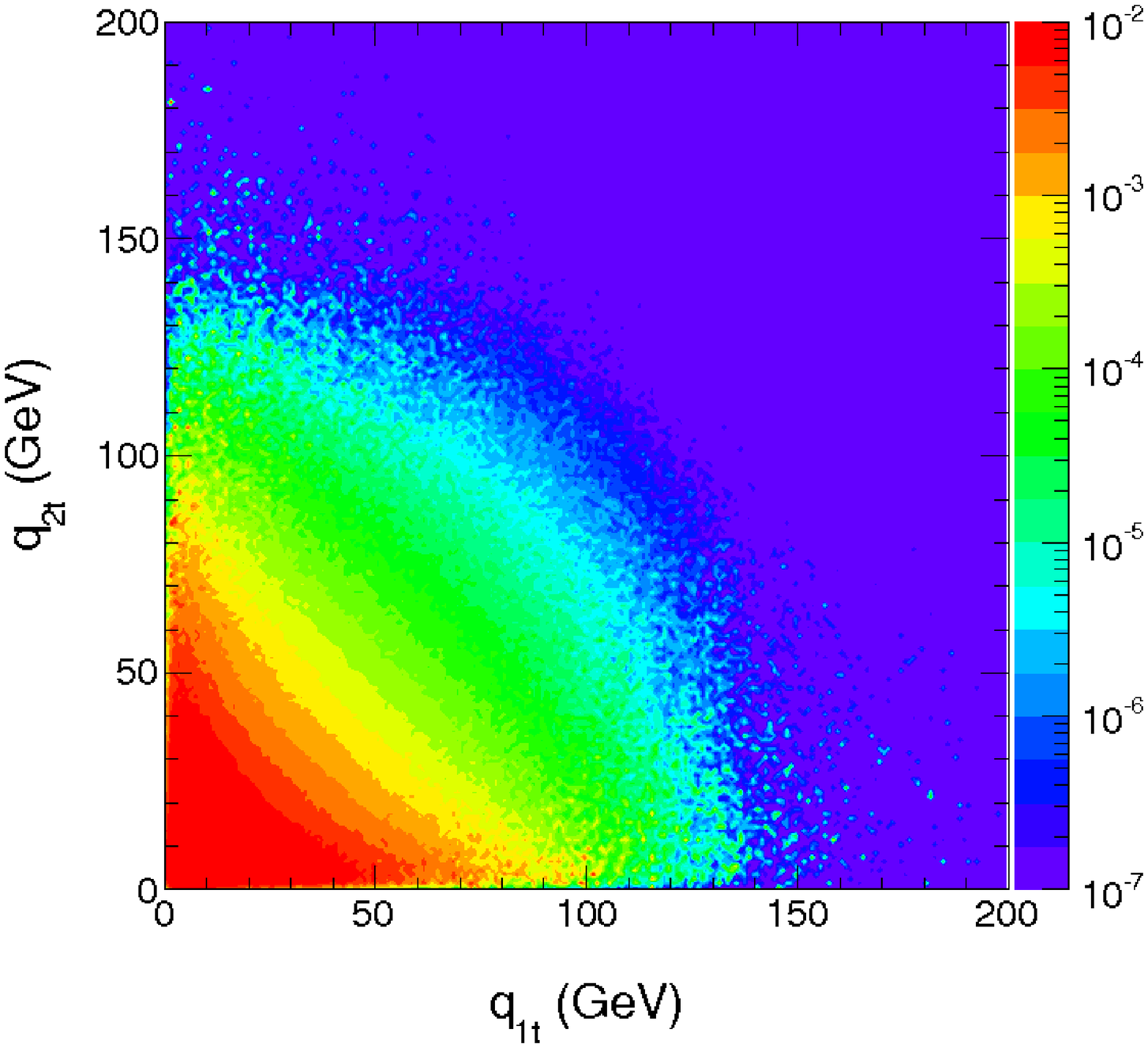}
\caption{
\small Two-dimensional distribution in $(q_{1t},q_{2t})$ for the 
KMR (left panel) and for the Jung CCFM (set$A0$) (right panel) UGDF and 
for $\mu^2 = m_{H}^2$.
}
\label{fig:dsig_dq1tdq2t_second_method}
\end{figure}

In Fig.~\ref{fig:dsig_dq1tdq2t_second_method} we show rather theoretical 
distributions in "initial" gluon transverse momenta. Those distributions
are almost identical to those discussed already for on-shell Higgs boson
production (see Fig.~\ref{fig:q1tq2t}). The distribution for the KMR UGDF
is broader than that for the Jung CCFM (set$A0$) UGDF.

\begin{figure}[!h]
\includegraphics[width=6cm]{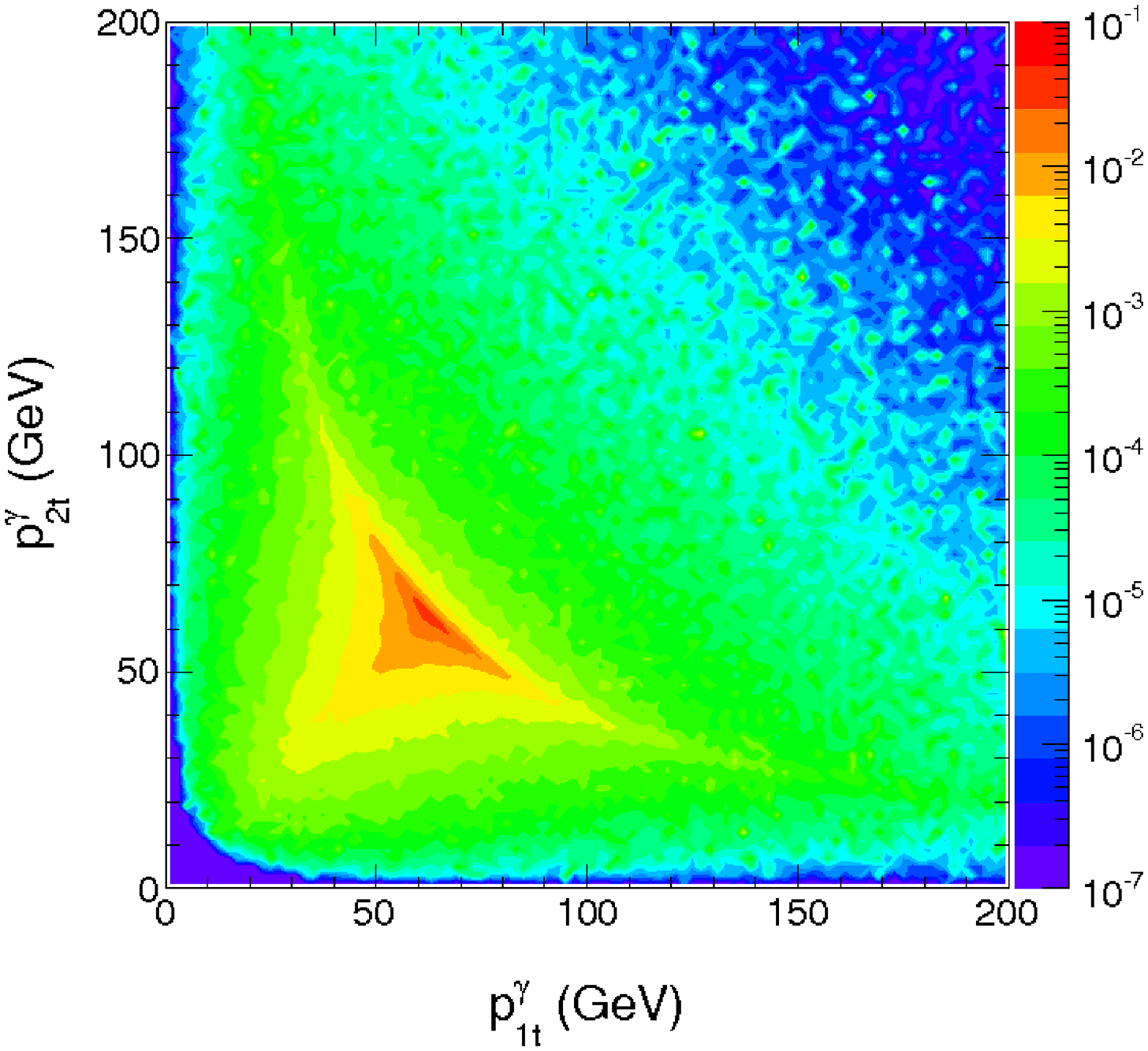}
\includegraphics[width=6cm]{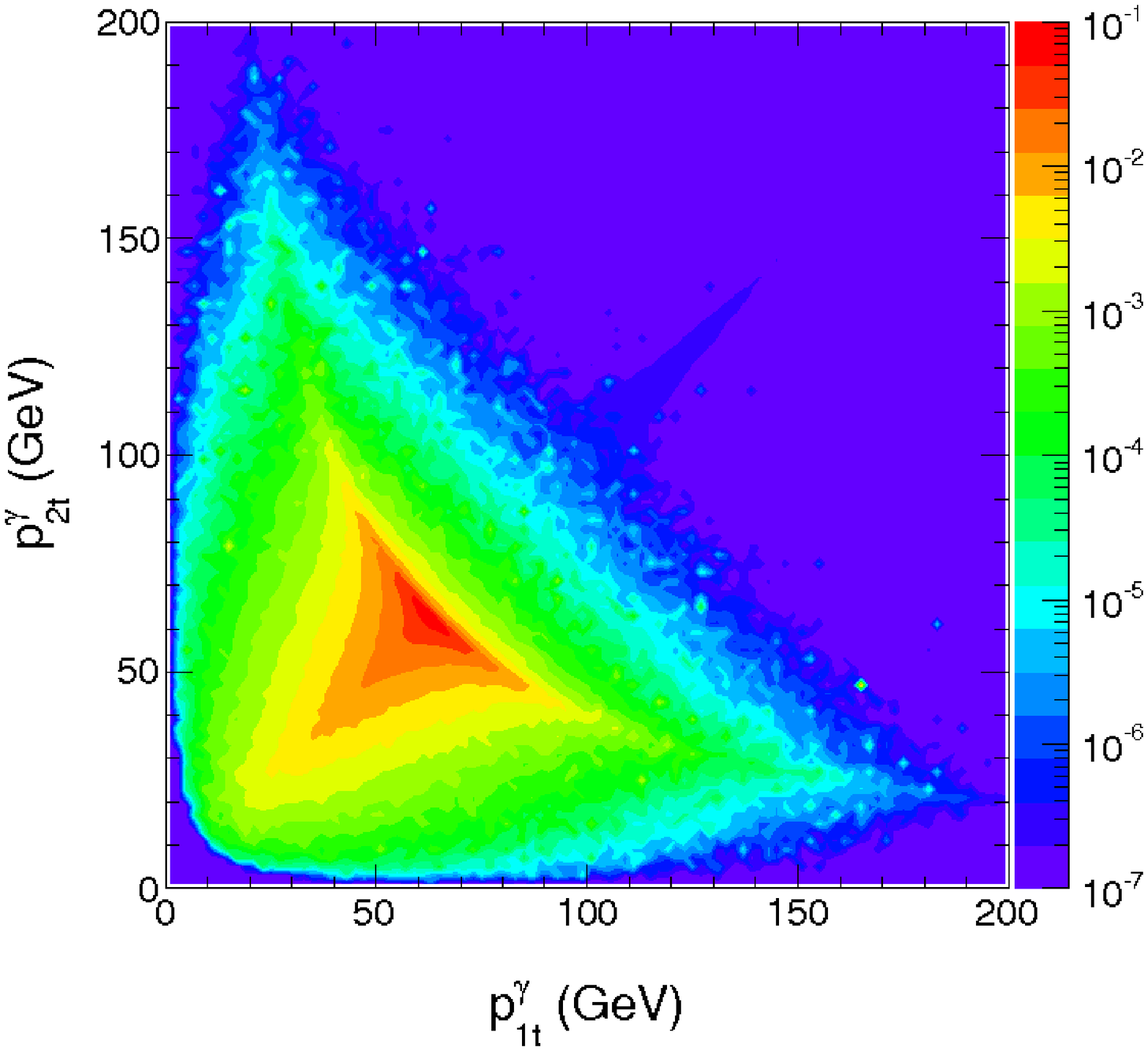}
\caption{
\small Two-dimensional distribution in photon transverse momenta 
$(p_{1t},p_{2t})$ for the KMR (left panel) and for the Jung CCFM
(set$A0$) (right panel) UGDF and for $\mu^2 = m_{H}^2$.
}
\label{fig:p1tp2t_second_method}
\end{figure}

Finally we wish to present two-dimensional correlations in photon
transverse momenta (see Fig.~\ref{fig:p1tp2t_second_method}).
Again this distribution is similar to its counterpart obtained within
first method (compare Fig.~\ref{fig:p1tp2t}).

\subsection{Higgs in association with one jet (gluon)}

Now we wish to show some results of calculation for 
$\textrm{Higgs} + \textrm{gluon}$ 
production within $k_t$-factorization approach.

\begin{figure}[!h]
\includegraphics[width=6cm]{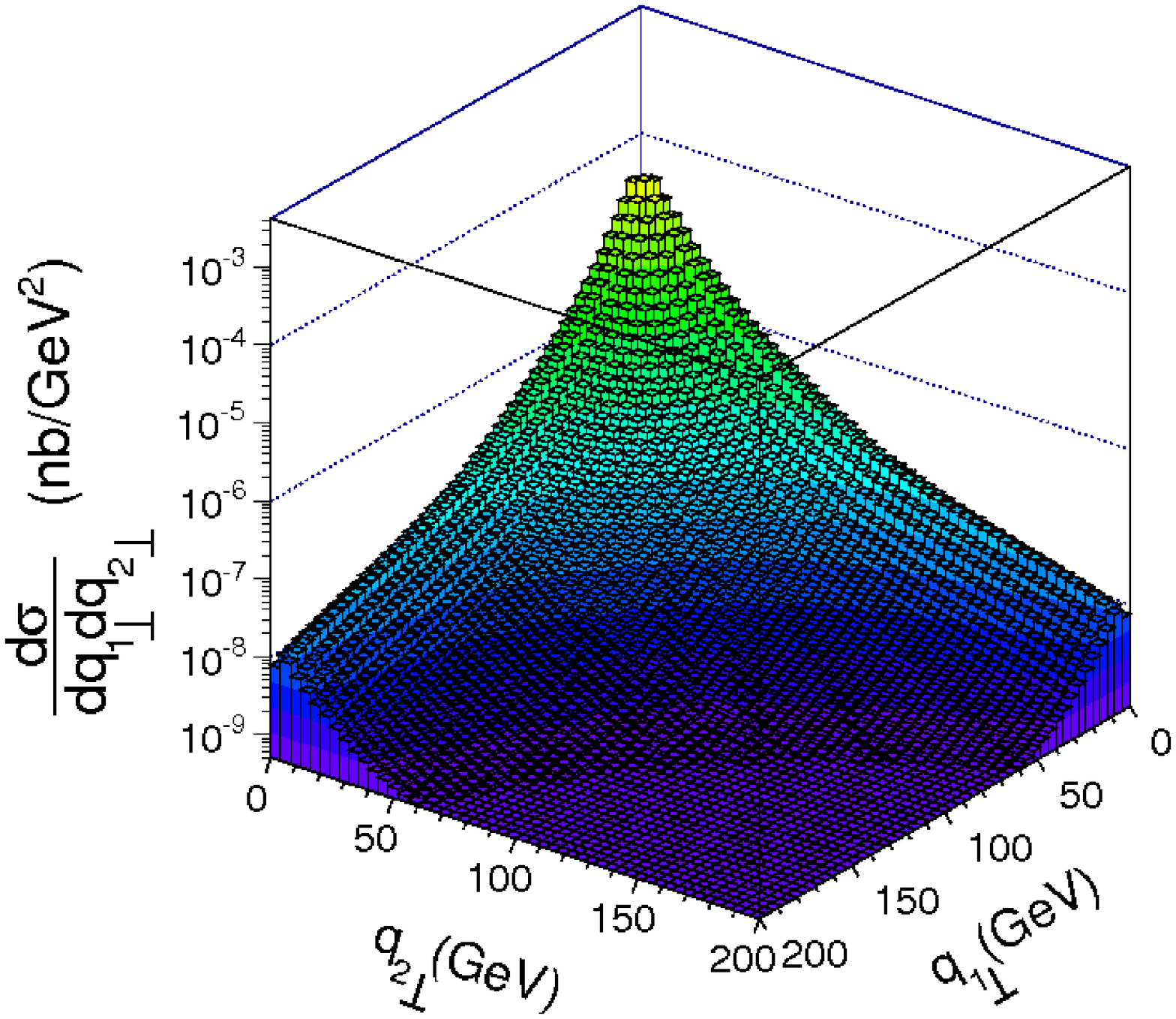}
\includegraphics[width=6cm]{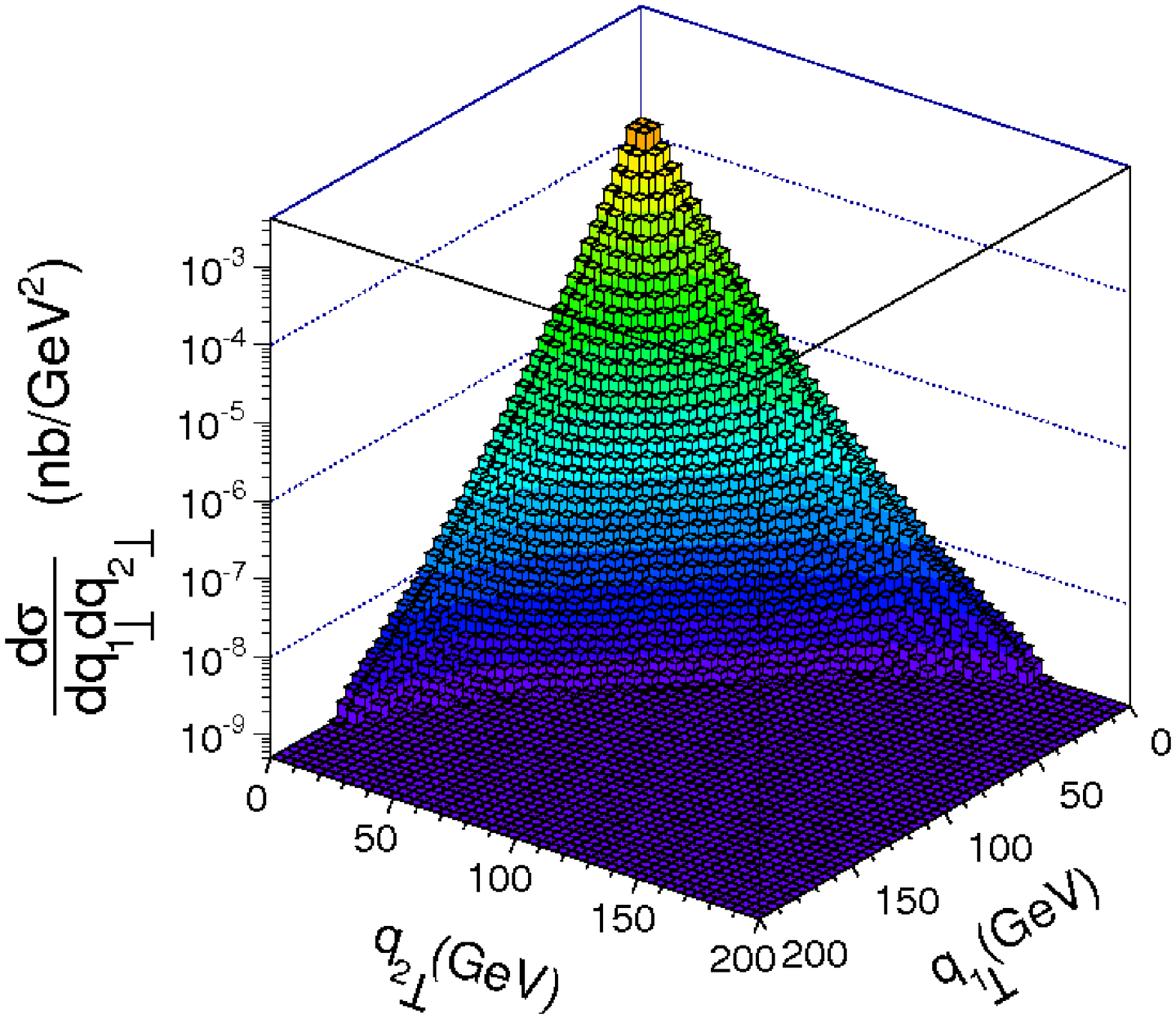} \\
\includegraphics[width=6cm]{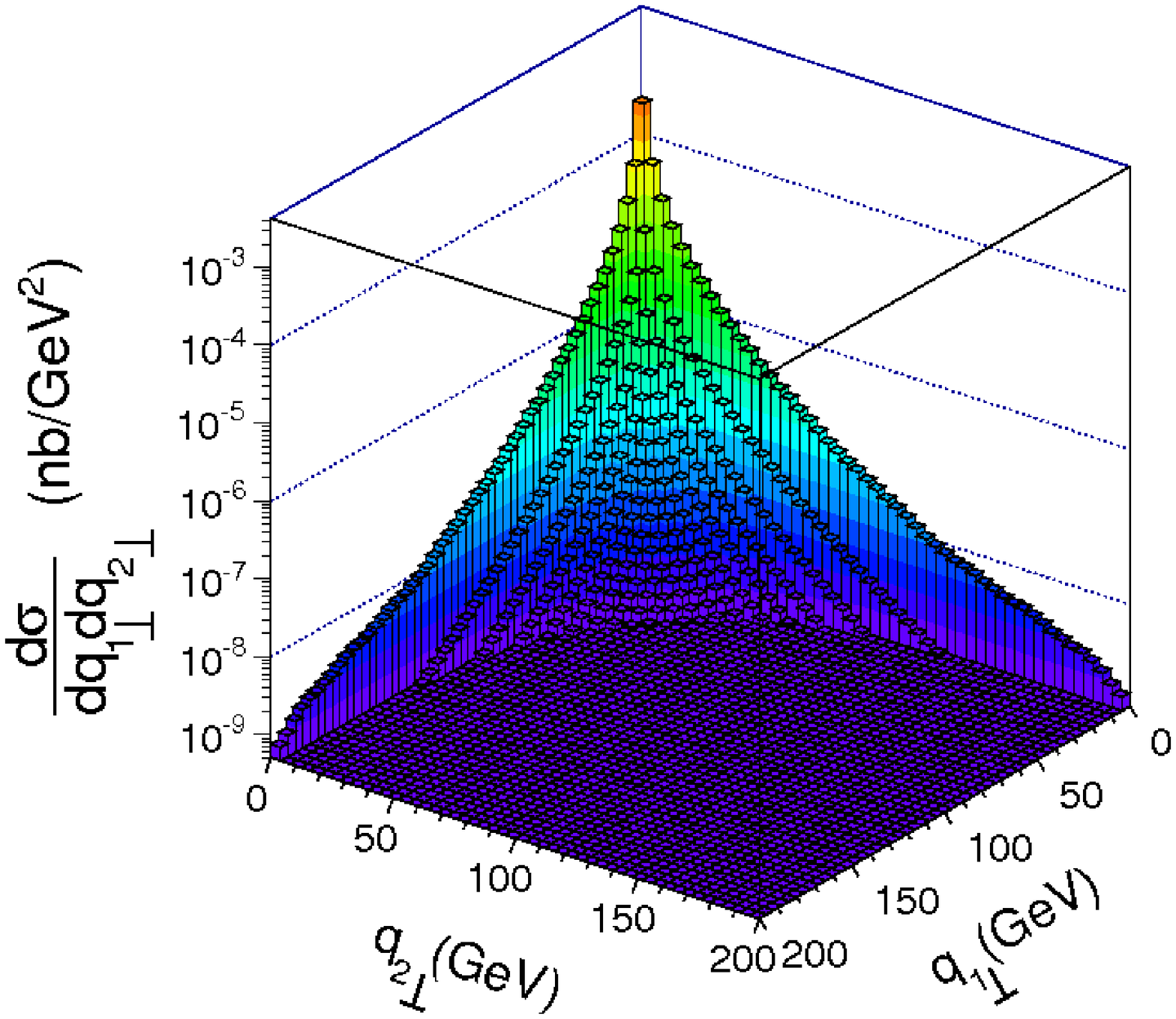}
\includegraphics[width=6cm]{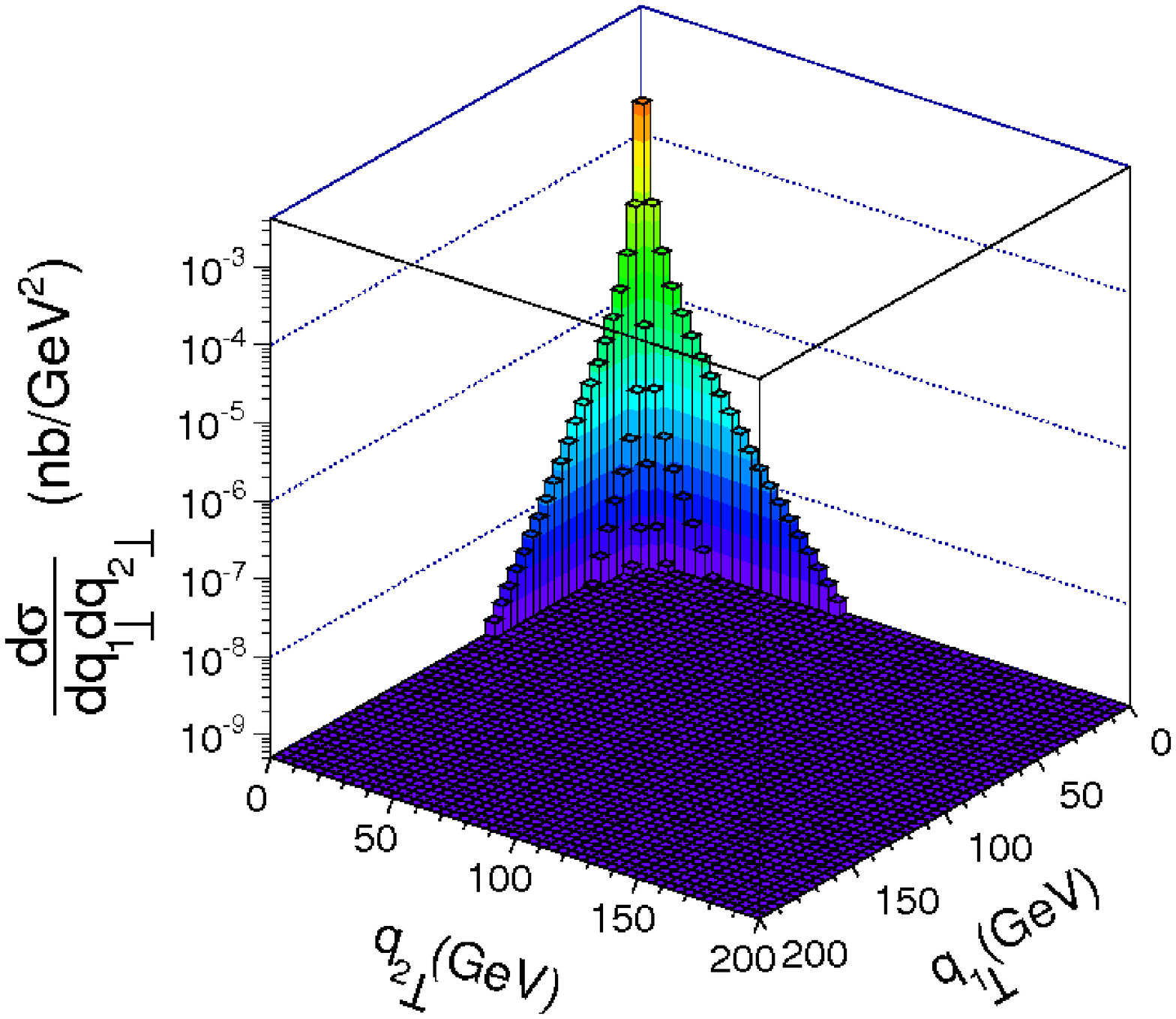} 
\caption{
\small Two-dimensional distribution in $(q_{1t},q_{2t})$ for the
$g g \to H g$ process for the four different UGDFs used previously
also for the $g g \to H$ calculation: 
KMR UGDF and $\mu_F^2 = m_{H}^2$ (left top panel)
and Jung CCFM set$A0$ (right top panel), Kutak-Sta{\'s}to (left bottom panel)
and Kutak-Sapeta (right bottom panel).
}
\label{fig:dsig_dq1tdq2t_gg_Hg}
\end{figure}

We start from a pedagogical two-dimensional distributions
(similar distribution was discussed in the context of the $g g \to H$ mechanism)
in initial gluon transverse momenta ($q_{1t}, q_{2t}$).
In Fig.~\ref{fig:dsig_dq1tdq2t_gg_Hg} we show distribution for the four
different UGDFs used also for the $g g \to H$ calculation.

\begin{figure}[!h]
\includegraphics[width=8cm]{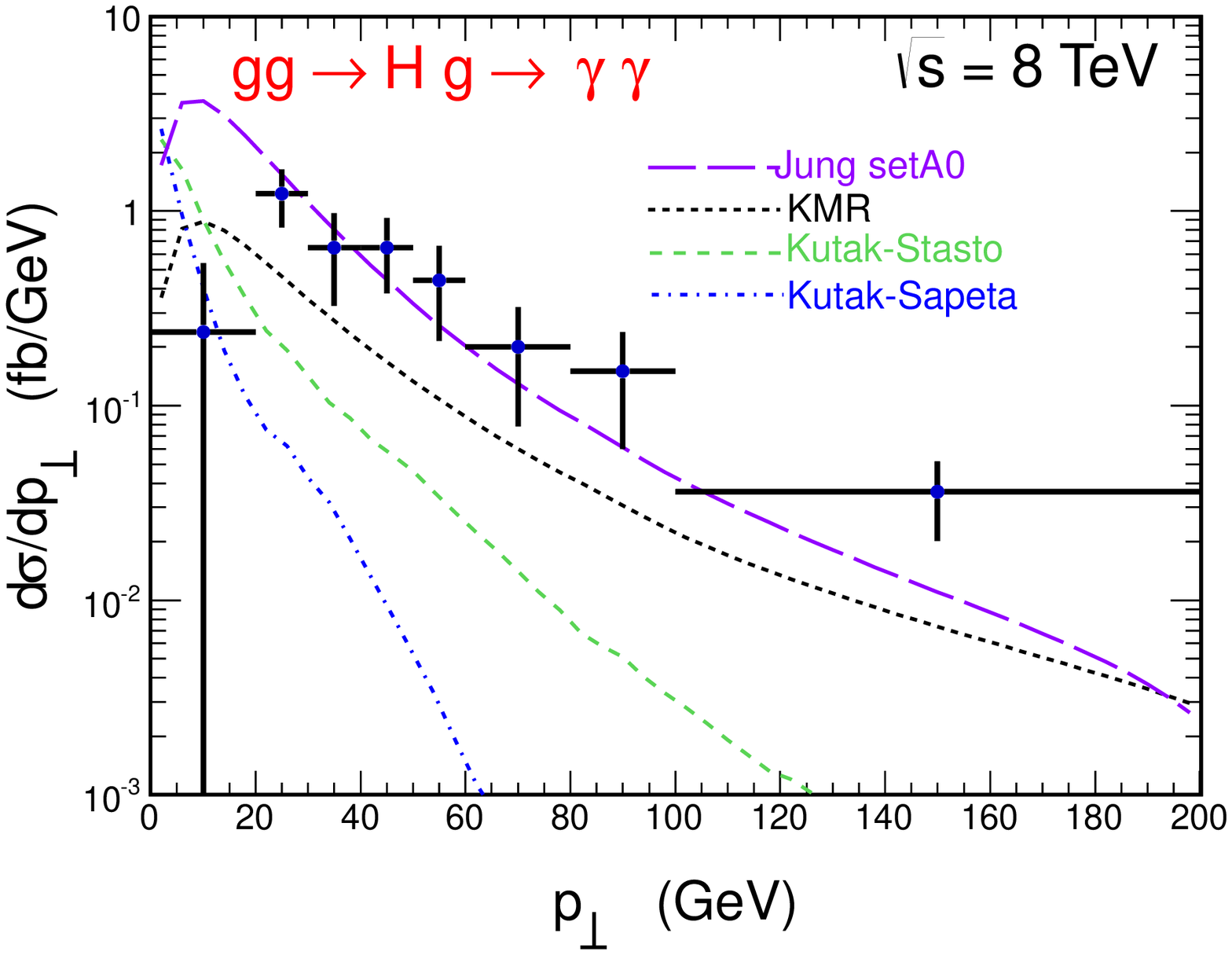}
   \caption{
\small Transverse momentum distribution of the Higgs boson in 
the $\gamma \gamma$-channel produced in the $g g \to H g$ subprocess 
for the different UGDFs from the literature.
}
 \label{fig:dsig_dpt_NLO}
\end{figure}

The Higgs transverse momentum distribution is particularly interesting
in the context of the preliminary ATLAS data.
In Fig.~\ref{fig:dsig_dpt_NLO} we show corresponding distributions
for the four UGDFs used in the present study.
It is worth to notice that the inclusion of gluon transverse momenta
automatically removes singular behaviour of the cross section at
$p_t \to$ 0. We observe that the cross section for $gg\to Hg$ is of the 
same order of magnitude as that calculated before for $gg \to H$. 
We wish to notice here that in contrast to other gluon initiated
processes the dominant piece of the $gg \to Hg$ is not included in 
the calculation of $gg \to H$. This can be easily understood by inspecting
diagrams in Fig.~\ref{fig:gg_H} and Fig.~\ref{fig:NLO}. While for 
the $gg \to H$ fusion the triangle with top quarks is the dominant
mechanism, in the case of the $gg \to Hg$ process these are 
the diagrams with top-quark boxes that dominate. Due to 
their completely different topology the diagrams with boxes are
certainly not contained in our previous calculations for the $gg \to H$ fusion. 
The same is true for all previous calculations
of the Higgs boson production in the $k_t$-factorization 
\cite{LZ2005,PTS2006,LS2006_Higgs,LMZ2014}. 

\begin{figure}[!h]
\includegraphics[width=8cm]{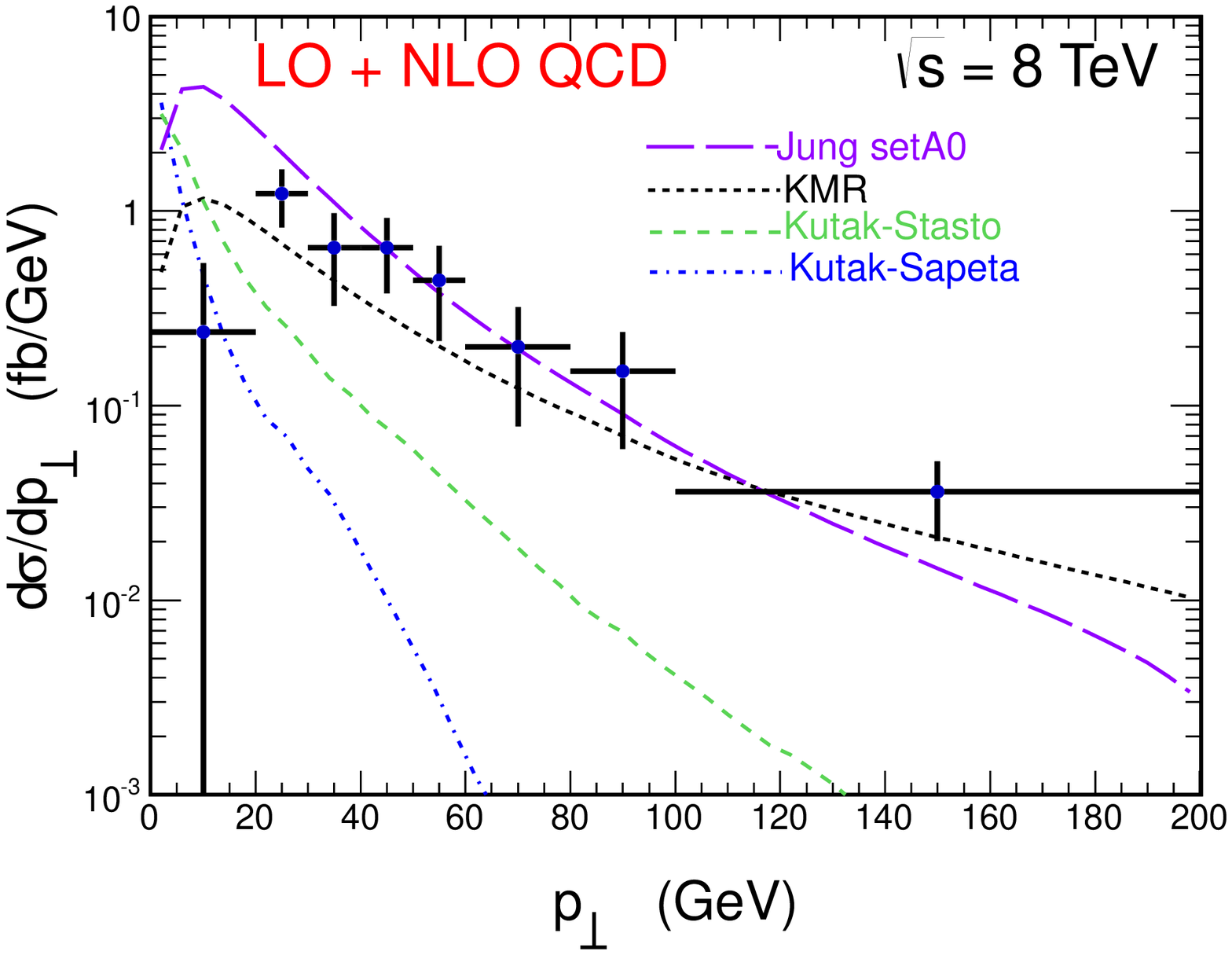}
   \caption{
\small Transverse momentum distribution of the Higgs boson in 
the $\gamma \gamma$-channel produced in the $g g \to H$ and in
the $g g \to H g$ subprocesses for the different UGDFs from the literature.
}
 \label{fig:dsig_dpt_LO_NLO}
\end{figure}

In Fig.~\ref{fig:dsig_dpt_LO_NLO} we show sum of the leading 
($g g \to H$) and the next-to-leading ($g g \to H g$) contributions
again for the different UGDFs used so far. The result for the KMR and
Jung CCFM set$A0$ UGDFs is already almost consistent with the new ATLAS data.
The electroweak contribution will be discussed below.

\subsection{Higgs in association of two jets}

\begin{figure}[!h]
\includegraphics[width=6cm]{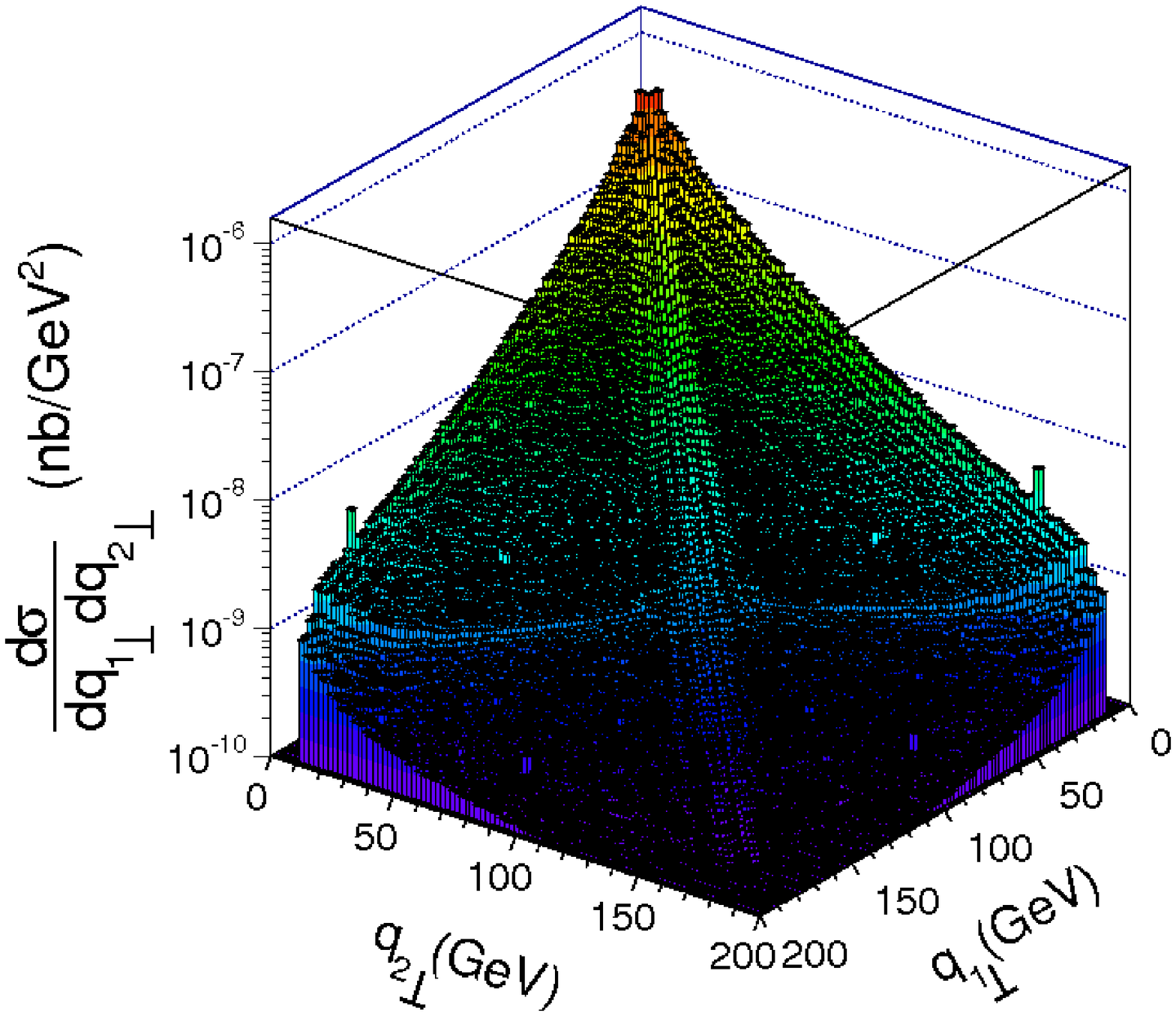}
\includegraphics[width=6cm]{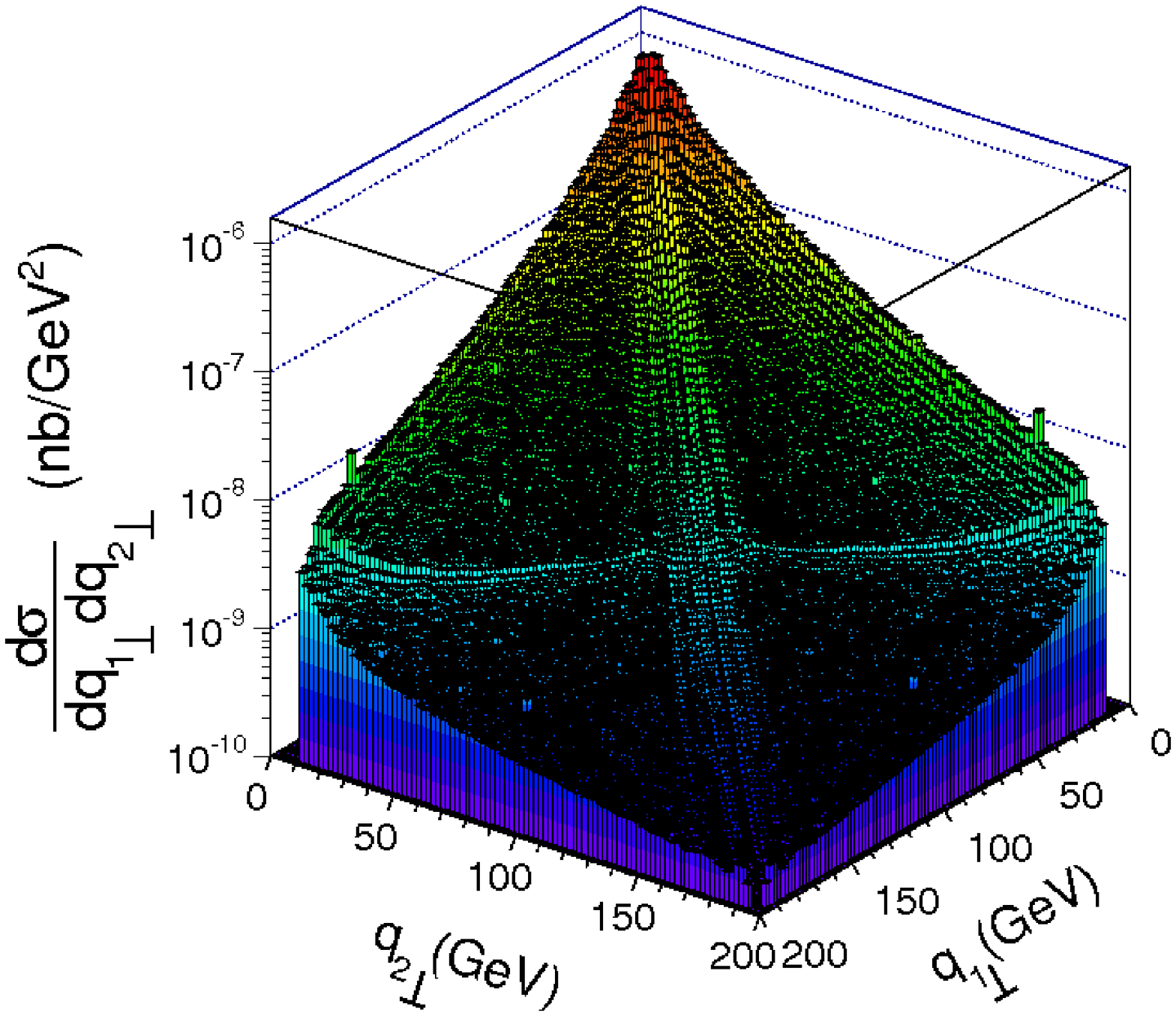}
\caption{
\small Two-dimensional distribution in jet transverse momenta 
$(p_{3t},p_{4t})$ for the $2 \to 3$ process $g g \to g H g$ (left)
and $i j \to i H j$ (right). In this calculation $\mu_F^2 = m_{H}^2$
and $\mu_{r,1}^2 = p_{3t}^2$, $\mu_{r,2}^2 = p_{4t}^2$.
A cut on $p_{3t},p_{4t} >$ 10 GeV has been assumed in addition. 
}
\label{fig:p3tp4t_gg_gHg}
\end{figure}

It is interesting to compare the $k_t$-factorization calculation
at large $q_{1t}$ and $q_{2t}$ (transverse momenta of the fusing gluons) 
with standard (collinear) calculation of the Higgs boson production associated 
with two jets. In Fig.~\ref{fig:p3tp4t_gg_gHg} we show two dimensional distribution
in the space of the transverse momenta of the associated jets
($p_{3t}$, $p_{4t}$) for the $g g \to g H g$ process only. 
Since initial gluons are collinear this is also exactly distribution 
in $(q_{1t},q_{2t})$ (transverse momenta of the t-channel gluons) 
and can be directly compared with similar distributions obtained 
previously in the $k_t$-factorization $gg \to H$ calculation. In this calculation
high-energy limit and rapidity ordering (see Ref.~\cite{Duca_high_energy}) 
was assumed. The shape here is similar to that for the KMR UGDF.
However, the absolute normalization is sizeably smaller.
We think that such contributions are therefore
effectively included in the calculation with the KMR UGDF.
But this is certainly not true for saturation-inspired UGDFs.

The contributions of the $g q({\bar q})$, $q({\bar q})$ and
$(q({\bar q}),q({\bar q})$ discussed previously in the formalism section
are usually not included explicitly in the $k_t$-factorization approach
with most of UGDFs (except of the KMR UGDF) and has to be taken into account 
when comparing theoretical results to experimental data.

\begin{figure}[!h]
\includegraphics[width=7cm]{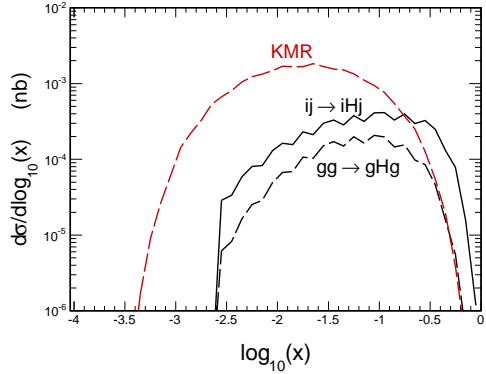}
\caption{
\small
$\log_{10}(x_i)$ distribution for the $k_t$-factorization approach for $gg \to H$ 
(upper dashed line, red online) and for 
$g g \to g H g$ (lower dashed line) and $i j \to i H j$ (solid line) 
for $q_{1t}, q_{2t} >$ 10 GeV.
}
\label{fig:x1orx2_kt_vs_gHg}
\end{figure}

Let us make further comparison of results for $g g \to g H g$ with 
$p_{3t}, p_{4t} <$ 10 GeV (which automatically means
$q_{1t}, q_{2t} <$ 10 GeV) with similar result obtained within
the $k_t$-factorization approach for $gg \to H$ with the KMR UGDF.
From Table \ref{table:Higgs_total_cross_section} we see that 
the result for the KMR UGDF is much bigger than that for the 
$g g \to g H g$ collinear-factorization approach.
This is difficult to understand as in the KMR model the whole
transverse momentum is generated in the last step of the ladder.
In Fig.~\ref{fig:x1orx2_kt_vs_gHg} we show distributions
in $\log_{10}(x_1)$ or $\log_{10}(x_2)$ for both cases.
One clearly sees that $x$'s for the $k_t$-factorization approach
(maximum at $\log_{10}(x_i) \approx$ -1) are
smaller than their counterparts for the $g g \to g H g$ (maximum at
$\log_{10}(x_i) \approx$ -2). This explains huge cross section
at large $q_{1t}$ and/or large $q_{2t}$ within the $k_t$-factorization
approach for $gg \to H$ which does not include fully correctly the kinematics of 
the actual process 
(missing jets are not included in calculating $x_1$ and $x_2$).
It is not clear to us how to consistently correct the calculation 
for the kinematical effect.

\subsection{Other contributions}

\begin{figure}[!h]
\includegraphics[width=8cm]{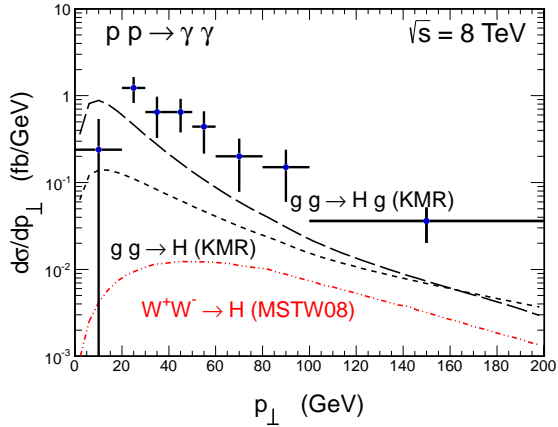}
   \caption{
\small Transverse momentum distribution of the Higgs boson
in the $\gamma \gamma$ channels for different 
mechanisms: $g g \to H$ (solid line), $g g \to H g$ (dashed line) and 
$W W \to H$ (dash-dotted line).
}
 \label{fig:pt_gamgam_mechanisms}
\end{figure}

In Fig.~\ref{fig:pt_gamgam_mechanisms} we compare contributions of
different mechanisms. The QCD contributions shown in this subsection 
were calculated with the KMR UGDF.
Surprisingly the contribution of the
next-to-leading order mechanism $g g \to H g$ is even slightly bigger than that
for the $g g \to H$ fusion, especially for intermediate Higgs boson transverse
momenta. As already discussed there is almost no double counting when
adding the corresponding cross sections due to quite different Feynman 
diagram topology.
As shown in the present analysis the $g g \to H$ mechanism is not sufficient 
within the $k_t$-factorization approach.
The $2 \to 3$ contribution of the $g g \to g H g$ subprocess is probably
also quite large but here one can expect that a big part is already 
contained in the $gg \to H$ calculation especially with the KMR UGDF. 
Therefore we do not add this contribution explicitly when
calculating $d\sigma/dp_{t,sum}$.
The contribution of the $WW$, $ZZ$ fusion is also fairly sizeable.
In principle, the Higgs bosons (or photons from the Higgs boson) could 
be to some extend isolated by requiring rapidity gap i.e. production 
of Higgs boson isolated off other hadronic activity.

If we added the contribution together we would almost describe the ATLAS
data.

In the future one could include into such an analysis
even higher order $g g \to g H g$ contribution as well as
associated production $g g \to t H \bar t$, $q \bar q \to WH$ and 
$q \bar q \to ZH$. Their contributions are known to be only slightly
smaller than the contribution of the $WW$ and $ZZ$ fusion.

\section{Conclusions}

In the light of new ATLAS data we have carefully analysed Higgs boson 
production in the $\gamma \gamma$ channel.
We have concentrated rather on QCD contributions. 
The $g g \to H$ mechanism has been considered within $k_t$-factorization
approach. Different unintegrated gluon distributions from the literature
have been used. In general, the cross section for the leading-order Higgs 
production within $k_t$-factorization approach is somewhat smaller than 
its counterpart for the leading-order collinear approximation.

We have calculated the cross section for $g g \to H \to \gamma \gamma$ 
within two methods.
In the first method we have performed decay of the on-shell Higgs
boson within a Monte Carlo method using the $H \to \gamma \gamma$ branching fraction known from 
the literature. In the second method we have performed direct
calculation with explicit $2 \to 2$ $g g \to H^* \to \gamma \gamma$ 
subprocess. In the second method the intermediate Higgs boson
is off-mass-shell. The two methods give slightly different results.
We have carefully discussed corresponding differences. The second, more 
proper method leads to a small enhancement of the cross section with respect to the first
method. If this is the explanation of the enhancement of the $\gamma \gamma$
channel as observed by the ATLAS and CMS collaborations requires 
further studies.

In contrast to recent claims in the literature, the leading-order
$g g \to H$ calculation does not describe the preliminary
ATLAS data when correct Standard Model couplings are taken
into account. Higher-order corrections within $k_t$-factorization
such as $g g \to H g$ have been discussed in addition. 
Their contribution turned out to be of similar order as that 
for $g g \to H$. We have argued that there is almost no double
counting when adding the leading-order $gg \to H$ and next-to-leading order
$gg \to Hg$ contributions in the $k_t$-factorization approach.
The reason is that the box diagrams dominate for the $gg \to Hg$ subprocess
and they are not present in the leading-order $gg \to H$ subprocess.  
Also $i j \to i H j$ ($i,j = q, \bar q g$) 
collinear NNLO contributions have been shown to be rather sizeable, 
also those with quarks and/or antiquarks
that are certainly not included in the leading-order $k_t$-factorization
approach.

In addition, we have calculated purely electroweak contributions of
the $WW$ and $ZZ$ fusion and associated production
$q q' \to W H$ and $q q \to Z H$. In general, the electroweak
contributions are also not negligible.

The sum of all (QCD and electroweak) contributions gives a result 
which is almost consistent with the ATLAS preliminary data. This requires, however, 
a further analysis as some double-counting between the leading 
($g g \to H$), next-to-leading ($g g \to H g$) and NNLO 
($g g \to g H g$) order contributions have to be carefully studied
in this approach.

In summary, the production of the Higgs boson in the $\gamma \gamma$
channel can be used to test unintegrated gluon distributions
provided all contributions to the cross section are
carefully taken into account.

\vspace{1cm}

{\bf Acknowledgments}
We are indebted to Simone Marzani for information about
some references related to our work and Nikolai Zotov 
for pointing out a  misprint in our first manuscript on arXiv.
This work was partially supported by the Polish NCN grants: 
DEC-2011/01/B/ST2/04535 and DEC-2013/09/D/ST2/03724 .



\begin{thebibliography}{100}

\bibitem{Higgs_discovery}
G.~Aad {\it et al.} (the ATLAS Collaboration), Phys. Lett. {\bf B716}, 1 (2012);\\
S.~Chatrchyan {\it et al.} (the CMS Collaboration), Phys. Lett. {\bf B716}, 30 (2012).

\bibitem{Aad:2013wqa} 
  G.~Aad {\it et al.}  (the ATLAS Collaboration),
  Phys.\ Lett.\ {\bf B726}, 88 (2013); corrigendum: Phys.\ Lett.\ {\bf B734}, 406 (2014).

\bibitem{Chatrchyan:2013mxa} 
  S.~Chatrchyan {\it et al.}  (the CMS Collaboration),
  Phys.\ Rev.\ {\bf D89}, 092007 (2014).

\bibitem{Khachatryan:2014iha} 
  V.~Khachatryan {\it et al.}  (the CMS Collaboration),
  arXiv:1405.3455 [hep-ex].

\bibitem{Aad:2014aba} 
  G.~Aad {\it et al.}  (the ATLAS Collaboration),
  arXiv:1406.3827 [hep-ex].

\bibitem{NNLO}
R.V. Harlander and W.B. Kilgore,
Phys. Rev. Lett. {\bf 88} (2002) 201801;\\
C. Anastasiou and K. Melnikov,
Nucl. Phys. {\bf B646} (2002) 220;\\
V. Ravindran, J. Smith and W.L. van Neerven,
Nucl. Phys. {\bf B665} (2003) 325.

\bibitem{FFGT2011}
D. de Florian, G. Ferrera, M. Grazzini and D. Tommasini,
JHEP {\bf 1111} (2011) 064, arXiv:1109.2109 [hep-ph].

\bibitem{FFGT2012}
D. de Florian, G. Ferrera, M. Grazzini and D. Tommasini,
JHEP {\bf 1206} (2012) 132, arXiv:1203.6321 [hep-ph].

\bibitem{Jung2013}
P. Cipriano, S. Dooling, A. Grebenyuk, P. Gunnellini, F. Hautmann,
H. Jung and P. Katsas,
Phys. Rev. {\bf D88} (2013) 097501.

\bibitem{LMZ2014}
A.V. Lipatov, M.A. Malyshev and N.P. Zotov, Phys. Lett. {\bf B735}, 79 (2014); arXiv:1402.6481 [hep-ph].

\bibitem{KMR}
M.A. Kimber, A.D. Martin and M.G. Ryskin, Phys. Rev. {\bf D63} (2001) 114027;\\
G. Watt, A.D. Martin and M.G. Ryskin, Eur. Phys. J. {\bf C31} (2003) 73.

\bibitem{Jung}
H. Jung, G.P. Salam, Eur. Phys. J. {\bf C19} (2001) 351;\\
H. Jung, arXiv:0411287 [hep-ph].

\bibitem{ATLAS_Higgs}
ATLAS collaboration, ATLAS note, ATLAS-CONF-2013-072.

\bibitem{Kutak-Stasto}
K. Kutak and A.M. Sta{\'s}to, Eur. Phys. J {\bf C41} (2005) 343.

\bibitem{Kutak-Sapeta}
K. Kutak and S. Sapeta, Phys. Rev. {\bf D86}, 094043 (2012); arXiv:1205.5035 [hep-ph].

\bibitem{ESW_book}
R.K. Ellis, W.J. Stirling and B.R. Webber,
``QCD and Collider Physics'', Cambridge University press,
Cambridge.

\bibitem{LS2006_Higgs}
M. {\L}uszczak and A. Szczurek, Eur. Phys. J. {\bf C46} (2006) 123.

\bibitem{Hautmann:2002tu}
 F.~Hautmann,
 Phys.\ Lett.\ B {\bf 535} (2002) 159
 [hep-ph/0203140].

\bibitem{KS2004}
J. Kwieci\'nski and A. Szczurek, Nucl. Phys. {\bf B680} (2004) 164.

\bibitem{LZ2005}
A.V. Lipatov and N.P. Zotov, Eur. Phys. J. {\bf C44} (2005) 559.

\bibitem{PTS2006}
R.S. Pasechnik, O.V. Teryaev and A. Szczurek,
Eur. Phys. J. {\bf C47} (2006) 429.

\bibitem{Marzani:2008az}
 S.~Marzani, R.~D.~Ball, V.~Del Duca, S.~Forte and A.~Vicini,
 Nucl.\ Phys.\ B {\bf 800} (2008) 127
 [arXiv:0801.2544 [hep-ph]].

\bibitem{Barger:1987nn} 
  V.~D.~Barger and R.~J.~N.~Phillips,
  Redwood City, USA: Addison-Wesley (1987) 592 P. (Frontiers in Physics, 71)

\bibitem{EGN1976}
J.R.~Ellis, M.K.~Gaillard, D.V.~Nanopoulos,
Nucl. Phys. B {\bf 106}, 292 (1976).

\bibitem{DHPRS2011}
A. Denner, S. Heinemeyer, I. Puljak, D. Rebuzzi and M. Spira,
Eur. Phys. J. {\bf C71}, 1753 (2011); arXiv:1107.5909 [hep-ph].

\bibitem{DKS98}
A. Djouadi, J. Kalinowski and M. Spira, Comput. Phys. Commun.
{\bf 108} (1998) 56.

\bibitem{PDG}
J. Beringer et al. (Particle Data Group),
Phys. Rev. {\bf D86} (2012) 010001.

\bibitem{LL1997}
Y. Liao and X. Li, Phys. Lett. {\bf B396} (1997) 225.

\bibitem{LS2006_ccbar}
M. {\L}uszczak and A. Szczurek, Phys. Rev. {\bf D73} (2006) 054028.

\bibitem{Lepage1978}
G.P.~Lepage, J. Comput. Phys. {\bf 27}, 192 (1978).

\bibitem{2to2_basis}
R.K. Ellis, I. Hinchliffe, M. Soldate and J.J. van der Bij,
Nucl. Phys. B297 (1988) 221;\\
U. Baur and E.W.N. Glover,
Nucl. Phys. {\bf B339} (1990) 38.

\bibitem{WW_NLO}
T. Figy, D. Zeppenfeld and C. Oleari,
Phys. Rev. {\bf D68} (2003) 073005.

\bibitem{CD84}
R.N. Cahn and S. Dawson,
Phys. Lett. {\bf B136} (1984) 196.

\bibitem{CDD08}
M. Ciccolini, A. Denner and S. Ditmaier,
Phys. Rev. {\bf D77} (2008) 013002.

\bibitem{BMMZ2010}
P. Bolzoni, F. Maltoni, S.-O. Moch and M. Zaro,
Phys. Rev. {\bf D85} (2012) 035002.

\bibitem{GNY78}
S.L. Glashow, D.V. Nanopolous and A. Yildis,
Phys. Rev. {\bf D18} (1978) 1724.

\bibitem{Duca_Hjj}
V. Del Duca, W. Kilgore, C. Oleari, C. Schmidt and D. Zeppenfeld,
Phys. Rev. Lett. {\bf 87} (2001) 122001;
Nucl. Phys. {\bf B616} (2001) 367.

\bibitem{Duca_high_energy}
V. Del Duca, W. Kilgore, C. Oleari, C.R. Schmidt, D. Zeppenfeld,
Phys. rev. {\bf D67} (2003) 073003.

\bibitem{Duca_numerical_studies}
V. Del Duca, G. Kl\"amke, M.L. Mangano, M. Moretti, F. Piccinini,
A.D. Polosa and D. Zeppenfeld,
JHEP {\bf 0610} (2006) 016.

\bibitem{BP_book}
V. Barone and E. Predazzi,
``High-Energy Particle Diffraction'',
Springer, Berlin.

\bibitem{WMR2004}
G. Watt, A.D. Martin and M.G. Ryskin, Phys. Rev. {\bf D70} (2004)
014012; erratum: ibid. {\bf 70} (2004) 079902.

\bibitem{JS2001}
H. Jung and G.P. Salam, Eur. Phys. J. {\bf C19} (2001) 351.

\bibitem{KMR2001}
M.A. Kimber, A.D. Martin and M.G. Ryskin,
Phys. Rev. {\bf D63} (2001) 114027.

\bibitem{PS} 
T. Pietrycki and A. Szczurek,
Phys. Rev. {\bf D75} (2007) 014023;
T. Pietrycki and A. Szczurek, 
Phys. Rev. {\bf D76} (2007) 034003.

\bibitem{MS2013}
R. Maciu{\l}a and A. Szczurek,
Phys. Rev. {\bf D87} (2013) 094022.

\end{thebibliography}
\end{document}